\documentclass[3p, authoryear]{elsarticle}
\usepackage{amsmath,amssymb,mathrsfs}
\usepackage{stmaryrd}
\usepackage{latexsym}
\usepackage{CJK}
\usepackage{graphicx}
\usepackage{indentfirst}
\usepackage{listings}
\usepackage{xcolor}
\usepackage{booktabs}
\usepackage{stmaryrd}
\usepackage{multirow}
\usepackage{verbatim}
\usepackage{makecell}
\usepackage{lineno,hyperref}
\usepackage{longtable}
\usepackage{hyperref}
\usepackage{upgreek}
\usepackage{mathtools}
\usepackage{enumitem}
\usepackage{float}
 
\hypersetup{colorlinks,breaklinks,
            linkcolor=blue,urlcolor=blue,anchorcolor=blue,citecolor=blue}

\newcommand{\ud}{\mathrm{d}}

\newcommand{\uc}{\mathrm{c}}

\newcommand{\txd}{\text{d}}

\newcommand{\txw}{\text{w}}

\newcommand{\txI}{\text{I}}

\newcommand{\rmd}{\mathrm{d}}
\newcommand{\usgn}{\mathrm{sgn}}

\begin{document}

\title{Mesoscale Description of Interface-Mediated Plasticity} 

\author[cityu]{Jinxin Yu}
\author[hku]{Alfonso H. W. Ngan}
\author[hku]{David J. Srolovitz\corref{cor2}}
\author[cityu]{Jian Han\corref{cor1}}

\address[cityu]{Department of Materials Science and Engineering, City University of Hong Kong, Hong Kong SAR, China}
\address[hku]{Department of Mechanical Engineering, The University of Hong Kong, Pokfulam Road, Hong Kong SAR, China}
\cortext[cor1]{jianhan@cityu.edu.hk}
\cortext[cor2]{srol@hku.hk}

\date{\today}

\begin{abstract}
Dislocation-interface interactions dictate the mechanical properties of polycrystalline materials through
dislocation absorption, emission and reflection and interface sliding.
We derive a mesoscale interface boundary condition to describe these, based on bicrystallography and Burgers vector reaction/conservation. 
The proposed interface boundary condition is built upon Burgers vector reaction kinetics and is applicable to any type of interfaces in crystalline materials with any number of slip systems.
This approach is applied to predict slip transfer for any crystalline interface and stress state; comparisons are made to widely-applied empirical methods.
The results are directly applicable to many existing dislocation plasticity simulation methods.
\end{abstract}
\maketitle

\section{Introduction}


A classical approach to tailoring the mechanical properties (strength/ductility) of materials is through the manipulation of microstructure. 
Many common classes of microstructure may be described as a spatial distribution of interfaces; e.g., grain boundaries (GBs) in a polycrystal, interfaces between precipitates and a matrix, or phase boundaries in a multi-phase system. 
The principle behind the modulation of mechanical properties through microstructure design lies on the interactions between lattice dislocations (carriers of plastic deformation) and  homo-/hetero-phase interfaces. 
Hall-Petch strengthening~\citep{hall1951deformation,petch1953cleavage} is a remarkable example; it is commonly described as the result of dislocation pileups at impenetrable grain boundaries. 
To achieve strengthening and toughening simultaneously, researchers have designed and synthesized a spectrum of heterostructured materials~\citep{zhu2022heterostructured}, such as nanotwinned structures, gradient structured materials, heterogeneous lamella structured materials, dual-phase alloys, etc.
One triumph of such a strategy was the understanding of hetero-deformation induced strengthening~\citep{zhu2019perspective}, an interface-mediated mechanism. 
One consequence of grain-level plasticity interacting with GBs is the activation of GB sliding which, in turn, alters plasticity within the grains. 
Interface sliding was observed during plastic deformation of  bicrystals~\citep{lihua2022tracking} and polycrystalline materials~\citep{linne2020effect} and during superplasticity~\citep{wei2003superplasticity}. 
The focus of this work is the quantitative prediction of the interaction of plasticity within grains with interfaces in a rigorous, quantitative manner that respects crystallography and loading conditions. 
This approach is applicable to all types of homo- (GBs, twins, ...) and hetero-phase interfaces.

\begin{figure}[t]
\centering
\includegraphics[width=0.9\linewidth]{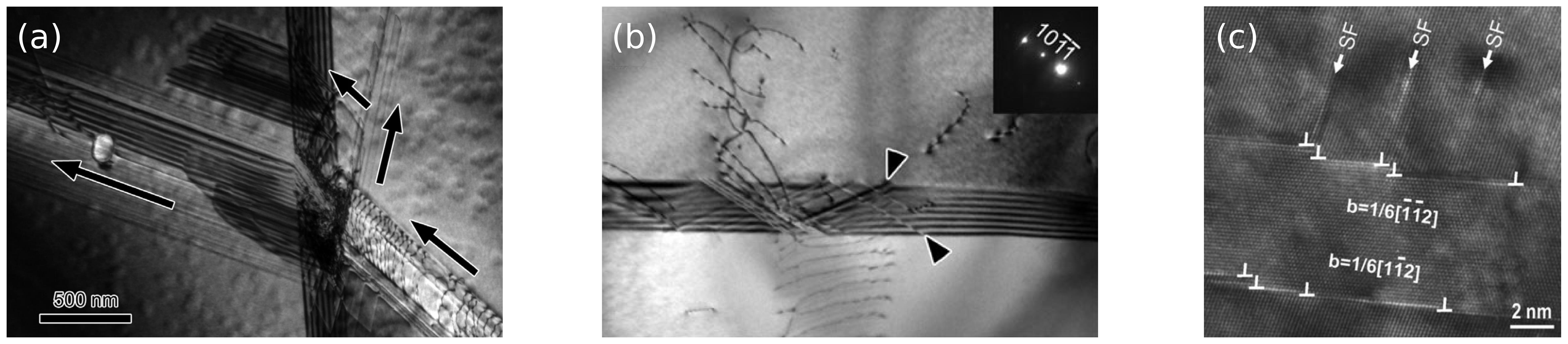}
\caption{\label{fig:tem}
Dislocation-GB interactions observed in TEM experiments. 
(a), (b), and (c) are reproduced with permission from, respectively, \citep{kacher2012quasi} (Copyright 2012 Elsevier), \citep{kacher2014situ} (Copyright 2014 Taylor \& Francis), and \citep{dao2006strength} (Copyright 2006, Elsevier). 
The arrows in (a) indicate the dislocation motion directions. 
The arrows in (b) indicate the GB dislocations. 
The GBs in (c) are coherent twin boundaries in Cu. 
}
\end{figure}

The interactions between lattice dislocations and interface have been widely explored in the literature since this interaction plays a major role in plasticity, strengthening, fracture, ...  
The classic picture of grain size/interface is that lattice dislocations pileup against interfaces, causing back stresses that reduce plastic deformation in grains; this is known as Hall-Petch strengthening.
Although highly simplified, this is a convenient starting point for thinking about the mechanisms by which dislocation/interface interactions affect plasticity.
There is considerable evidence that this pileup may be partially relaxed by slip transfer across an interface (i.e., precipitate cutting~\citep{gleiter1965beobachtung}). 
Transmission electron microscopy (TEM) observations have shown many different forms of  dislocation-interface interactions; even within the same material. 
Kacher et al.~\citep{kacher2012quasi} showed that dislocations may be  ``transmitted'' across a GB and/or be ``reflected'' back into their source-grain (see Fig.~\ref{fig:tem}a). 
Other observations show that  GBs/interfaces themselves may evolve as a result of plastic deformation; e.g., dislocation-GB interactions  lead to the formation~\citep{dao2006strength} and motion~\citep{ng2009deformation,linne2020effect} of  line defects within the GB/interface plane (see Fig.~\ref{fig:tem}b and c). 
These phenomena have also been observed in atomistic and multi-scale simulations~\citep{jin2006interaction,jin2008interactions,terentyev2018grain,tsuru2009fundamental,dewald2006multiscale,dewald2007multiscale,dewald2011multiscale}. 
Hence, GBs/interfaces are much more than simply blocking dislocations; they are a mediator of plasticity in the grains -- sliding, transmitting, absorbing and reflecting lattice dislocations, ...  
And, GBs/interfaces in a single material system can behave very differently, depending on bicrystallography and loading.


Extensive experimental observations of dislocation/GB interactions led Lee, Robertson, Birnbaum, and co-workers~\citep{lee1989prediction,lee1990tem,lee1990situ,kacher2014dislocation} to propose several criteria to identify the factors that account for slip transfer  across a particular GB for a particular loading condition (i.e., the ``LRB'' criteria). 
The  LRB criteria suggest that slip transfer is favorable for the slip systems in two grains which 
(i) are well aligned (in both slip planes/directions),  
(ii) result in the smallest residual Burgers vector left in the GB, and 
(iii) possess large Schmid factors. 
While these criteria are based upon sound reasoning, they are empirical and  not always mutually consistent. 
Atomistic~\citep{de2003modeling,bachurin2010dislocation,sangid2011energy,tsuru2016predictive,wang2008atomistic} and  coupled atomistic/discrete dislocation methods~\citep{dewald2006multiscale,kvashin2021interaction} were  employed to examine the LRB criteria. 
Dewald and Curtin~\citep{dewald2006multiscale,dewald2007multiscale,dewald2011multiscale}  modified the LRB criteria to account for dislocation dissociation, non-Schmid stresses and step formation. 
Nonetheless, the proposed criteria simply represent a systemization of observations. 
A bicrystallographically-based, theoretical understanding should not only be consistent with TEM observation and be consistent with mechanistic observations from experiments and atomistic simulations, but should be predictive. 


Several mesoscale, dislocation-based simulation models have been proposed for interface-mediated plastic deformation. 
Commonly used dislocation-based mesoscale, single crystal, plasticity  models include discrete dislocation dynamics (DDD)~\citep{lesar2020advances}, continuum dislocation dynamics (CDD)~\citep{acharya2001model,hochrainer2007three,leung2015new} and crystal plasticity finite element methods (CPFEM)~\citep{roters2010overview}.
Although Neumann boundary condition (BC) oversimplifies what happens at interfaces, this approach is easily applied; hence, in most mesoscale simulations, GBs are treated as Neumann BCs (i.e., impenetrable to dislocations or plastic strain); this BC was employed in  CPFEM~\citep{dunne2007lengthscale,evers2004scale,jiang2022grain}, DDD~\citep{jiang2019effects} and CDD~\citep{stricker2016slip} simulations. 
Note that even if an interface is treated as a Neumann BC,  plastic deformation on one side of the interface affects that on the other side (through a  dislocation pileup induced concentrated stress field). 
Neumann BCs do not account for the deformation induced evolution of an interface, interface sliding nor does it distinguish between interfaces of different types -- except (rarely) through changes of empirical parameters.



Some simulation approaches model the effects of dislocation-interface interactions by invoking the empirical LRB criteria~\citep{cho2020dislocation,zhang2021dislocation,quek2014polycrystal}. 
For example, in one case, dislocation transmission through a GB is assumed to occur through a Frank-Read mechanism when criterion (iii) is satisfied~\citep{zhou2012dislocation}.
In others, both criteria (ii) and (iii) were considered to determine whether a dislocation will penetrate a GB~\citep{fan2012toward}.
In some cases, LRB-like criteria (i.e.,  criteria established upon slip system alignment and the resolved shear stress) were invoked to determine dislocation absorption/emission  at a GB~\citep{zhang2021dislocation}.
Other semi-empirical GB models were incorporated into CPFEM to study  slip transfer in  polycrystals~\citep{cermelli2002geometrically,ma2006studying,mayeur2015incorporating}; 
e.g.,  Ma et al.~\citep{ma2006studying} considered dislocation transmission  according to criterion (ii) (i.e., the minimum residual Burgers) in a CPFEM study. 
Here, the GB is modeled by an element where the slip resistance is assumed to be proportional to the square of the minimum residual Burger. 
Mayeur et al.~\citep{mayeur2015incorporating}  modeled a GB as an interface affected zone in CPFEM, where the probability of each transfer event is determined based upon criteria (i) and (iii). 
While these models  achieved some success, the empirical nature of the LRB criteria makes such models unreliable. 
Also note that these models did not include the possibility of dislocation glide in the GB/GB sliding. 


Many phenomenological models include  interactions between dislocations and GBs within a strain gradient plasticity framework~\citep{fredriksson2007modelling,gottschalk2016computational,erdle2017gradient}.
A general GB model was developed by Gurtin~\citep{gurtin2008theory} based on the continuum thermodynamic Coleman-Noll procedure, which was partially implemented in a finite element method by \"{O}zdemir et al.~\citep{ozdemir2014modeling} (this approach couples  bulk  and GBs via a microscopic force balance).
One merit of this method is that grain misorientation and GB orientation may be included through slip-interaction moduli.
This method was extended to a microstructure-motivated higher-order internal boundary conditions and numerically implemented to investigate the influence of GBs on shear deformation, at both macroscopic and microscopic levels~\citep{van2013grain}.
Piao and Le~\citep{piao2022thermodynamic} noted that  standard strain gradient plasticity models  ignore the configurational entropy and effective temperature which enables dislocations to satisfy  universal laws for plastic flows~\citep{le2020two}; an alternative continuum approach that accounts for configurational entropy was developed and applied to study the dislocation transmission/absorption at a GB at a continuum level.
While these models are theoretically more rigorous, the physical mechanisms observed in experiments and atomistic simulations are not explicitly reflected. 
For example,  GB/interface dislocations (residual dislocations) are not arbitrary; their Burgers vectors must correspond to the bicrystallographic translational symmetry~\citep{pond1977absorption,king1980effects,balluffi1982csl,han2018grain}.  
Again, these methods do not correctly capture  interface sliding induced by dislocation-interface interactions, although widely observed in  experiments and atomistic simulations~\citep{dao2006strength,ng2009deformation,linne2020effect,ZHU202042};
Gurtin~\citep{gurtin2008theory} did propose a GB sliding approach, but this has not been implemented (to our knowledge). 


In this paper, we  address the same question  raised by Gurtin~\citep{gurtin2008theory}: \textit{``Is there a physically natural method of characterizing the possible interactions between the slip systems of two grains that meet at a grain boundary -- a method that could form the basis for the formulation of grain boundary conditions?''}.
More general, this question applies to all interfaces -- not just GBs. 
Our approach is based upon irreversible thermodynamics (Onsager relations~\citep{onsager1931reciprocal} and Ziegler's maximum entropy production principle~\citep{ziegler2012introduction}). 
We first propose a Burgers vector reaction-based dislocation-interface interaction model motivated by a wide range of experimental observations and consistent with crystallography constraints.
The admissible Burgers vectors of the lattice dislocations are determined by the orientation and lattice structure of the adjoining grains. 
The admissible Burgers vectors of interface dislocations (or disconnections) must be consistent with the displacement-shift-complete (DSC) vectors determined from the bicrystallography~\citep{pond1977absorption,king1980effects,balluffi1982csl,han2018grain}.
We describe  Burgers vector reactions by a linear relationship between  dislocation fluxes entering, leaving and within the interface and the applicable driving forces. 
The kinetic coefficient tensor is constrained by the requirement of Burgers vector conservation (or equivalently, compatibility of interface deformation). 
The overall kinetics are described by a tensorial Robin boundary condition at the interface that is consistent with the linear response theory. 
Our model is bicrystallography-respecting, explicit about Burgers vector reactions, and simple to implement within different plasticity simulation schemes.


The paper is organized as follows. 
In Section~\ref{theory}, a general model for individual dislocation reactions at interface is formulated as an interface  boundary condition and then extended to multiple slip systems/reactions case (several special cases are discussed to validate the interface BC).
In Section~\ref{applications}, the mesoscale interface BC is applied to a one-dimensional continuum dislocation dynamics (CDD) model to examine the dislocation-GB interactions under different crystallographic orientations.
We then propose rigorous dislocation transmission conditions and compare those with the predictions from  LRB criteria. the transmitted dislocation density is defined to measure the amount of dislocations reacted at the interface and then they are compared with each LRB criterion. 
The results show that while the  LRB criteria are reasonable, they fail in many cases and are not always self-consistent.


\section{Interface boundary condition}\label{theory}

\begin{figure}[t]
\centering
\includegraphics[width=0.9\linewidth]{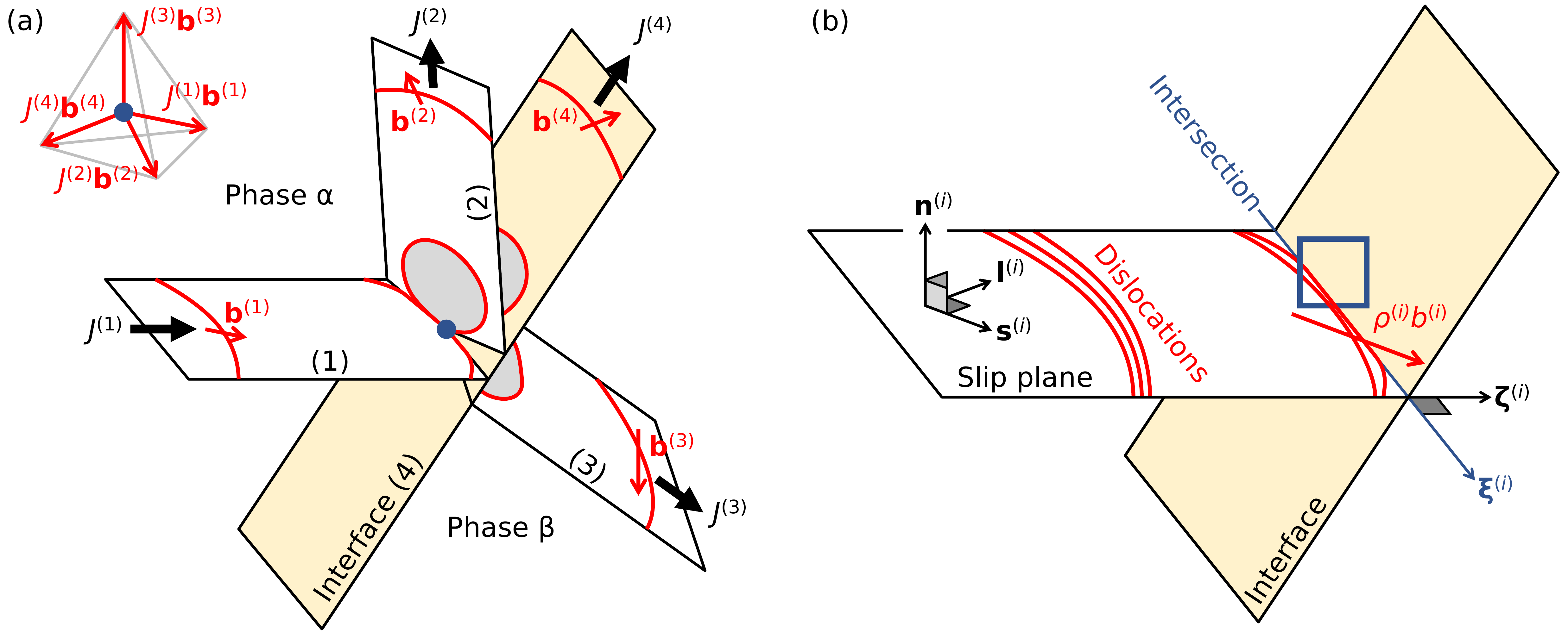}
\caption{\label{fig:bc_3d}
(a) Schematic of a dislocation reaction when the dislocations on slip plane (1) glide to the interface (4). 
$J^{(i)}$ and $\mathbf{b}^{(i)}$ denote the dislocation flux and Burgers vector on plane $(i)$. 
The reaction products $\mathbf{b}^{(2)}$, $\mathbf{b}^{(3)}$ and $\mathbf{b}^{(4)}$ are related to  dislocation reflection, dislocation transmission and GB sliding, respectively. 
Since the total Burgers vector is conserved, the four vectors, $\{J^{(i)}\mathbf{b}^{(i)}\}$ ($i=1,2,3,4$) form a tetrahedron with the origin located at the centroid, as shown in the inset. 
(b) Configuration of the interaction between a slip plane and the interface. 
The blue square at the intersection represents a unit area threaded by dislocations.
}
\end{figure}

Consider the interface-slip system configuration  illustrated in Fig.~\ref{fig:bc_3d}a, which shows two slip planes in Phase $\alpha$ (labeled (1) and (2)) and a single slip plane in Phase $\beta$ (labeled (3)); the two phases  are divided by an interface plane (labeled (4)). 
The Burgers vector of the dislocations on plane $(i)$ is $\mathbf{b}^{(i)}$. 
$\mathbf{b}^{(1)}$, $\mathbf{b}^{(2)}$ and $\mathbf{b}^{(3)}$ are  lattice dislocation Burgers vectors.
$\mathbf{b}^{(4)}$, on the other hand, is not an arbitrary vector in the interface (like the residual Burgers vector commonly defined in the literature), but must be a DSC lattice vector (i.e., determined from the bicrystallography)~\citep{kegg1973grain,pond1977absorption,SHEN19883231,elkajbaji1988interactions,lee1990situ,jin2006interaction,jin2008interactions,kacher2012quasi,tsuru2009fundamental,sangid2011energy,king1980effects,balluffi1982csl,han2018grain}. 
For example, for a coherent twin boundary in a face-centered cubic crystal, an admissible interface dislocation is a twinning partial with $\mathbf{b} = \langle 112\rangle a_0/6$ (see the dislocations along the twin boundaries in Fig.~\ref{fig:tem}c).

Suppose that  dislocations with $\mathbf{b}^{(1)}$ glide from the Phase $\alpha$ interior to the interface with a flux  $J^{(1)}$. 
The following Burgers vector reaction may occur where $\mathbf{b}^{(1)}$ contacts the interface (the blue point in Fig.~\ref{fig:bc_3d}a):
$J^{(1)}\mathbf{b}^{(1)}
\rightarrow
-J^{(2)}\mathbf{b}^{(2)}
-J^{(3)}\mathbf{b}^{(3)}
-J^{(4)}\mathbf{b}^{(4)}$,  
i.e., the dislocations $\mathbf{b}^{(2)}$, $\mathbf{b}^{(3)}$ and $\mathbf{b}^{(4)}$ are nucleated at the contact (the minus sign before $J^{(i)}$ denotes dislocations flowing away from the contact). 
In this case, $J^{(2)}$, $J^{(3)}$ and $J^{(4)}$ correspond to dislocation reflection, dislocation transmission and interface sliding, respectively. 
In general, reactions follow from Burgers vector (flux) conservation:
\begin{equation}\label{Burgersreaction}
\sum_{i=1}^4 J^{(i)}\mathbf{b}^{(i)} = \mathbf{0},
\end{equation}
i.e., the four vectors $\{J^{(i)}\mathbf{b}^{(i)}\}$ form a tetrahedron with the origin located at the centroid (see the Fig.~\ref{fig:bc_3d}a inset). 
Note that to keep dimensions consistent, we consider the flux along the interface  $J^{(4)}$ as the number of dislocations crossing a point in the interface (in 2D) per unit time divided by the interfacial width $\delta^\text{I}$.

Equation~\eqref{Burgersreaction} is not to be confused with a condition of zero net strain rate at the interface (this follows from how the fluxes $\{J^{(i)}\}$ are defined -- all pointing away from the contact point). For example, if two phases are identical with no interface (i.e., a single crystal),  dislocations should simply move across the contact point without  reaction;  in this case, $b^{(1)}=b^{(2)}$ and $J^{(1)}=-J^{(2)}$ such that Eq.~\eqref{Burgersreaction} is trivially satisfied and the strain strain rate is $J^{(1)}b^{(1)}=-J^{(2)}b^{(2)}$  (i.e., the results are compatible with, but not governed by, Eq.~\eqref{Burgersreaction}).
As a second example, consider an impenetrable, non-reflecting and static interface ($J^{(2)}=J^{(3)}=J^{(4)}=0$); here Eq.~\eqref{Burgersreaction} yields $J^{(1)}=0$, which is the classical Neumann BC.

We now focus on the slip plane $(i)$/interface intersection; for the example shown in Fig.~\ref{fig:bc_3d}b, the slip plane has unit normal  $\mathbf{n}^{(i)}$,   dislocation slip direction   $\mathbf{s}^{(i)}$ (parallel to $\mathbf{b}^{(i)}$),  and  line direction  $\boldsymbol{\xi}^{(i)}$, from which we define $\mathbf{l}^{(i)} \equiv \mathbf{n}^{(i)} \times \mathbf{s}^{(i)}$ and  $\boldsymbol{\zeta}^{(i)} \equiv \boldsymbol{\xi}^{(i)} \times \mathbf{n}^{(i)}$. 
At the dislocation/interface intersection, $\boldsymbol{\xi}^{(i)}$ is parallel to the intersection line and $(\mathbf{n}^{(i)}, \boldsymbol{\xi}^{(i)}, \boldsymbol{\zeta}^{(i)})$ form a local coordinate frame;  the dislocation density $\rho^{(i)}$ is the number of dislocation lines threading a unit area with normal $\boldsymbol{\xi}^{(i)}$. 
The dislocation flux is $\mathbf{J}^{(i)} = J^{(i)}\boldsymbol{\zeta}^{(i)}$, where $J^{(i)}$ is the number of dislocations cutting through a unit segment in $\mathbf{n}^{(i)}$ with velocity in $\boldsymbol{\zeta}^{(i)}$ per unit time. 
The Orowan equation defines the dislocation flux $J^{(i)} = \rho^{(i)}v^{(i)} = \dot{\gamma}^{(i)}/b^{(i)}$ , where $v^{(i)}$ is the dislocation velocity along the $\boldsymbol{\zeta}^{(i)}$-axis and $\dot{\gamma}^{(i)}$ is the shear rate on the $i^\textrm{th}$ slip system (the dislocation density on the interface is normalized by the interface width $\delta^\text{I}$). 
Hence,  Burgers vector conservation, Eq.~\eqref{Burgersreaction}, is equivalent to the requirement that $\sum_i \dot{\gamma}^{(i)} \mathbf{s}^{(i)} = \mathbf{0}$; this guarantees the continuity of the displacement rate component normal to the interface plane. 

The reaction kinetics described in Fig.~\ref{fig:bc_3d}b and Eq.~\eqref{Burgersreaction} may be developed based upon  linear response theory~\citep{onsager1931reciprocal}. 
We assume that, at a point on an interface where reaction occurs, the Burgers vector flux is linearly proportional to the driving force such that the interface BC is  
\begin{equation}\label{dissipation2}
\left(\begin{array}{c}
J^{(1)} b^{(1)} \\ 
J^{(2)} b^{(2)} \\ 
J^{(3)} b^{(3)} \\ 
J^{(4)} b^{(4)}
\end{array}\right)_{\txI}
=
\left(\begin{array}{cccc}
L_{11} & L_{12} & L_{13} & L_{14}\\
       & L_{22} & L_{23} & L_{24}\\
       &   & L_{33} & L_{34} \\
\text{sym.} & & & L_{44}
\end{array}\right)
\left(\begin{array}{c}
f^{(1)} \\ f^{(2)} \\ f^{(3)} \\ f^{(4)}
\end{array}\right)_{\txI}, 
\end{equation}
where the left-hand side represents a list of Burgers vector fluxes (not dislocation number fluxes), $f^{(i)}$ is the driving force exerted on  Burgers vector $\mathbf{b}^{(i)}$, and  the kinetic coefficient tensor  $\mathbf{L} = [L_{ij}]$ is  symmetric (in accordance with the Onsager reciprocity theorem). 
The subscript ``I'' denotes  quantities evaluated at a point on the interface. 
The derivation of Eq.~\eqref{dissipation2} based on the maximum entropy production principle~\citep{ziegler2012introduction,martyushev2006maximum} is given in \ref{linearresponse} -- in the derivation we assume local equilibrium and small driving forces. 

We may understand Eq.~\eqref{dissipation2} from another viewpoint.
When a dislocation glides in a grain interior, the dislocation velocity is often expressed by  $v=L\tau^m$, where $L$ is the mobility, $\tau$ is the resolved shear stress, and $m$ is unity (in the overdamped regime).
This equation can be rewritten as $Jb=Lf$, where $J=\rho v$ is the flux, and $f=\tau \rho b$ is the driving force. 
We assume the same form for the reaction kinetics at the interface in Fig.~\ref{fig:bc_3d}a, but with two differences.
First, the fluxes $J$ considered in Eq.~\eqref{Burgersreaction} correspond to the annihilation and production rates of dislocations at the interface, rather than movement of dislocations.
Hence, mobility $L$ should be treated as a reaction kinetic constant for Burgers vector annihilation and production, rather than dislocation mobility. 
(Once dislocations are produced at the interface from Burgers vector reaction, they move with different mobility laws within grains.)
Second, letting the flux $J^{(i)}$ on each slip system only depend on the driving force $f^{(i)}$ on that slip system alone; i.e., writing $J^{(i)}b^{(i)}=L^{(i)}f^{(i)}$ would not satisfy Eq.~\eqref{Burgersreaction} for general driving forces on different slip systems.
Instead, Eq.~\eqref{Burgersreaction} will be satisfied for general driving forces if $J^{(i)}$ also depends on driving force on other slip systems.
Therefore, the kinetics law governing the interfacial reaction is written in  tensorial form, Eq.~\eqref{dissipation2}; we now discuss each term in Eq.~\eqref{dissipation2}.


The driving forces on the dislocations, $\{f^{(i)}\}$, are Peach-Koehler (PK) force (see \ref{linearresponse}). 
On the $i^\text{th}$ slip system and under  stress $\boldsymbol{\sigma}$ (including external and internal stresses), the PK force is
\begin{equation}\label{PKforce}
f^{(i)}
= \mathbf{f}^{(i)} \cdot \boldsymbol{\zeta}^{(i)}
= \left[\boldsymbol{\sigma}(\rho^{(i)} \mathbf{b}^{(i)})\right] \times \boldsymbol{\xi}^{(i)} 
\cdot \boldsymbol{\zeta}^{(i)}
= \tau^{(i)} \rho^{(i)} b^{(i)}, 
\end{equation}
where $\boldsymbol{\xi}^{(i)}$ is the dislocation line direction at the interface, $\{\boldsymbol{\zeta}^{(i)},\boldsymbol{\xi}^{(i)},\mathbf{n}^{(i)}\}$ form a local coordinate frame as shown in Fig.~\ref{fig:bc_3d}b, and $\tau^{(i)} \equiv \mathbf{s}^{(i)}\cdot\boldsymbol{\sigma} \mathbf{n}^{(i)}$ is the resolved shear stress (RSS) on slip system ($i$) ($\mathbf{s}^{(i)}$ is the slip direction). 
Substituting the driving forces Eq.~\eqref{PKforce} into Eq.~\eqref{dissipation2}, the interface BC described by Eq.~\eqref{dissipation2} has the form of a Robin BC, i.e.,  flux is proportional to the densities $\{\rho^{(i)}\}$.

The dislocation fluxes at the interface are the result of dislocation reactions, which are constrained by Burgers vector conservation Eq.~\eqref{Burgersreaction}. 
This constraint requires that the coefficient tensor  takes the form 
\begin{align}\label{Lmatrix}
\mathbf{L}
&= \kappa
\left(\begin{array}{cccc}
(c^{(234)})^2 &
c^{(234)}c^{(314)} &
c^{(234)}c^{(124)} &
c^{(234)}c^{(132)}
\\
 &
(c^{(314)})^2 &
c^{(314)}c^{(124)} & c^{(314)}c^{(132)}
\\
 &
 &
(c^{(124)})^2 &
c^{(124)}c^{(132)}
\\
\text{sym.} & & & (c^{(132)})^2
\end{array}\right)= \kappa\mathbf{c}\otimes\mathbf{c},
\end{align}
where $\mathbf{c} \equiv (c^{(234)}, c^{(314)}, c^{(124)}, c^{(132)})^\text{T}$, $c^{(ijk)} \equiv \mathbf{s}^{(i)}\cdot\mathbf{s}^{(j)}\times\mathbf{s}^{(k)}$, and $\kappa$ is the overall reaction constant for reaction Eq.~\eqref{Burgersreaction} ($\kappa$ is discussed below). 
Hence, there is one kinetic parameter and all other parameters are purely geometric, which is reasonable as there is only one Burgers vector reaction involving four slip systems (including the slip systems in the interface). 
Clearly, Eq.~\eqref{Lmatrix} does guarantee that Eq.~\eqref{Burgersreaction} is always satisfied under  arbitrary driving forces.


We obtain the interface BC by substituting Eq.~\eqref{PKforce} and Eq.~\eqref{Lmatrix} into Eq.~\eqref{dissipation2}: 
\begin{equation}\label{dissipation4}
\mathbf{J}_\txI
=
\kappa \mathbf{B}^{-1} (\mathbf{c}\otimes\mathbf{c}) \mathbf{B} \mathbf{T}_\txI 
\boldsymbol{\rho}_\txI,
\end{equation}
where $\mathbf{J} \equiv (J^{(1)}, J^{(2)}, J^{(3)}, J^{(4)})^\text{T}$ and
$\boldsymbol{\rho} \equiv (\rho^{(1)}, \rho^{(2)}, \rho^{(3)}, \rho^{(4)})^\text{T}$ are the generalized dislocation flux and density, $\mathbf{B} \equiv \mathrm{diag}(b^{(1)}, b^{(2)}, b^{(3)}, b^{(4)})$, $\mathbf{T} \equiv \mathrm{diag}(\tau^{(1)}, \tau^{(2)}, \tau^{(3)}, \tau^{(4)})$, and the subscript ``I'' denotes the quantities evaluated at a point on the interface. 
In this equation,
$\mathbf{c} \otimes \mathbf{c}$ depends on how the slip systems in the two phases and the interface plane are oriented; hence, it may be related to  LRB criterion (i). 
$\mathbf{T}$ is simply a matrix of the resolved shear stresses, which may be related to the LRB Schmid factor criterion (ii). 
$\mathbf{B}$ is a matrix of the Burgers vectors, which may be related to the LRB  residual Burgers vector criterion (iii). 
In this sense,  interface BC Eq.~\eqref{dissipation4}, is  related to the three empirical LRB criteria; we  demonstrate this point in Section~\ref{applications}. 
$\kappa$ is the reaction constant which depends on the microscopic dislocation reaction mechanism, materials properties and temperature. 
Harmonic transition state theory suggests that it is reasonable to write 
$\kappa = (\kappa_0/T) e^{-Q/k_\text{B} T}$, 
where $T$ is the temperature, $k_\text{B}$ is the Boltzmann constant, $\kappa_0$ is an attempt frequency, and $Q$ is the energy barrier along the reaction path. 
The parameter $Q$ contains all the complexity at atomic scale: e.g., the reaction rate will depend sensitively on whether the dislocations are dissociated and  the stacking fault energy~\citep{dewald2006multiscale}. 
$Q$ may be determined from  atomistic simulations; e.g., Zhu et al.~\citep{zhu2007interfacial}  obtained  $Q$ for the interaction between a screw dislocation and a coherent twin boundary in copper by atomistic simulations. 
Their work also suggests that $Q$ may be extracted from experiments by, for example, measuring the the strain-rate sensitivity as a function of  stress.


Although the interface BC proposed for the reaction involving four slip systems (including the slip systems in the interface) is always valid, we examine  two special cases of the interface BC to show that it  behaves as expected. \\
\emph{Case 1: Two collinear Burgers vectors:}
e.g., $b^{(1)}\mathbf{s}^{(1)}$ and $b^{(2)}\mathbf{s}^{(2)}$  are  parallel (i.e., $\mathbf{s}^{(1)} = \mathbf{s}^{(2)}$) such that Eq.~\eqref{Burgersreaction} reduces to 
\begin{align}\label{interface_discuss3}
(J^{(1)}b^{(1)}+J^{(2)}b^{(2)})\mathbf{s}^{(1)}+J^{(3)}b^{(3)}\mathbf{s}^{(3)}+J^{(4)}b^{(4)}\mathbf{s}^{(4)}=\mathbf{0}.
\end{align}
If $\mathbf{s}^{(3)}$ and $\mathbf{s}^{(4)}$ are not collinear, we recover: $J^{(3)}b^{(3)}=J^{(4)}b^{(4)}=0$ and $J^{(1)}b^{(1)}=-J^{(2)}b^{(2)}$.
If $\mathbf{s}^{(3)}$ and $\mathbf{s}^{(4)}$ are also collinear  $J^{(1)}b^{(1)}=-J^{(2)}b^{(2)}$ and $J^{(3)}b^{(3)}=-J^{(4)}b^{(4)}$. 
In each case, the reaction only involves two slip systems. 
For example, based on Fig.~\ref{fig:bc_3d}a, direct transmission (without reflection and a residual Burgers vector) can occur when $\mathbf{b}^{(1)}$, $\mathbf{b}^{(3)}$ and the intersection between plane (1) and (3) are collinear.  \\
\emph{Case 2: Three co-planar Burgers vectors:}\label{two_dimen_case}
e.g.,   $b^{(1)}\mathbf{s}^{( 1)}$, $b^{(2)}\mathbf{s}^{(2)}$ and $b^{(3)}\mathbf{s}^{(3)}$, such that only two are independent. 
Choosing $\mathbf{s}^{(1)}$ and $\mathbf{s}^{(2)}$ as the basis vectors, $\mathbf{s}^{(3)}$ can be represented as a linear combination: $p\mathbf{s}^{(1)}+q\mathbf{s}^{(2)}$ ($p$ and $q$ are  combination coefficients).
Thus, Eq.~\eqref{Burgersreaction} becomes 
\begin{align}\label{interface_discuss2}
&(J^{(1)}b^{(1)}+pJ^{(3)}b^{(3)})\mathbf{s}^{(1)}
+ (J^{(2)}b^{(2)}+qJ^{(3)}b^{(3)})\mathbf{s}^{(2)}+J^{(4)}b^{(4)}\mathbf{s}^{(4)}=\mathbf{0};
\end{align}
i.e.,  three homogeneous equations in three unknowns (i.e., the coefficients before $\{\mathbf{s}^{(i)}\}$). 
The solution is $J^{(4)}b^{(4)} = 0$ and $J^{(1)}b^{(1)} : J^{(2)}b^{(2)} : J^{(3)}b^{(3)} = p:q:-1$. 
This reaction only involves  three slip systems with coplanar Burgers vectors. 
Since $p$ and $q$ are only geometry-dependent, the magnitudes of $\{J^{(i)}b^{(i)}\}$ ($i=1,2,3$) have only one degree of freedom. 
Case 2 applies in  two dimensions (2D); e.g., dislocations in monolayer graphene.  
In a 2D space, the vectors $\mathbf{J}$ and $\boldsymbol{\rho}$ contain three components, $\mathbf{B}$ and $\mathbf{T}$ are $3\times 3$, and $\mathbf{c} = (\mathbf{s}^{(2)} \cdot \mathbf{n}^{(3)}, \mathbf{s}^{(3)} \cdot  \mathbf{n}^{(1)}, \mathbf{s}^{(1)} \cdot \mathbf{n}^{(2)})^\text{T}$. 
One example in Section~\ref{applications} is an application of this 2D model.



Above, we only considered the case of a single Burgers vector reaction which may involve up to four slip systems (including  the interface). 
In practice, there may be multiple slip systems in each phase and the interface; reaction may occur amongst any quadruple of these.  
If there are $N$ slip systems, there will be $C_N^4 \equiv M$ possible reactions. 
We  label the $l^\text{th}$ reaction by  subscript ``$l$'' ($l = 1, \cdots, M$) such that Eq.~\eqref{dissipation4} (for  the $l^\text{th}$ reaction) is 
\begin{equation}\label{dissipation_multi}
\mathbf{J}_{l,\txI}
=
\kappa_l 
\mathbf{B}_l^{-1} 
(\mathbf{c}\otimes\mathbf{c})_l 
\mathbf{B}_l 
\mathbf{T}_{l,\txI} 
\boldsymbol{\rho}_{l,\txI},
\end{equation}
where all quantities with subscript $l$ are evaluated with  parameters for the four relevant slip systems, and $\kappa_l$ is the associated reaction constant. 
There are $M$ equations similar to Eq.~\eqref{dissipation_multi}. 
The total Burgers vector fluxes, due to all $M$ reactions is (see \ref{multi} for details)
\begin{equation}\label{rbc10}
\bar{\mathbf{J}}_\txI
= \sum_{l=1}^M \bar{\mathbf{J}}_{l,\txI}
= \bar{\mathbf{B}}^{-1}
\left[
\sum_{l=1}^M \kappa_l
\overline{(\mathbf{c}\otimes\mathbf{c})_l}
\right]
\bar{\mathbf{B}}
\bar{\mathbf{T}}_\txI
\bar{\boldsymbol{\rho}}_\txI,
\end{equation}
where the overline indicates the extended quantities: 
$\bar{\mathbf{J}} \equiv (J^{(1)}, \cdots, J^{(N)})^\text{T}$, 
$\bar{\boldsymbol{\rho}} \equiv (\rho^{(1)}, \cdots, \rho^{(N)})^\text{T}$, 
$\bar{\mathbf{B}} \equiv \mathrm{diag}(b^{(1)}, \cdots, b^{(N)})$, 
$\bar{\mathbf{T}} \equiv \mathrm{diag}(\tau^{(1)}, \cdots, \tau^{(N)})$, 
and $\overline{(\mathbf{c}\otimes\mathbf{c})_l}$ is the matrix $(\mathbf{c}\otimes\mathbf{c})_l$ extended to $N\times N$ dimensions with zero padding. 
The general result, Eq.~\eqref{rbc10}, is the major result of this paper. 
The reaction constants $\{\kappa_l\}$ are associated with the detailed atomic-scale  mechanisms for the $M$ reactions and thus, depend on  interface  structure. 
In this sense, $\{\kappa_l\}$ should also depend on the macroscopic degrees of freedom of an interface  (misorientation, inclination, misfit, ...).

\section{Applications}\label{applications}

The interface BC, Eq.~\eqref{dissipation4} or \eqref{rbc10}, may be easily implemented in different simulation methods, such as CDD, DDD and CPFEM. 
While we have done such implementations, detailed descriptions are beyond the scope of this paper. 
Below, we  demonstrate the application of the interface BC  for the  case of a minimal, one-dimensional (1D) CDD model for simplicity and transparency. 
The goal here is to examine how the interface BC works and if the result is consistent with the empirical LRB criteria. 

\begin{figure}[t]
\centering
\includegraphics[width=0.84\linewidth]{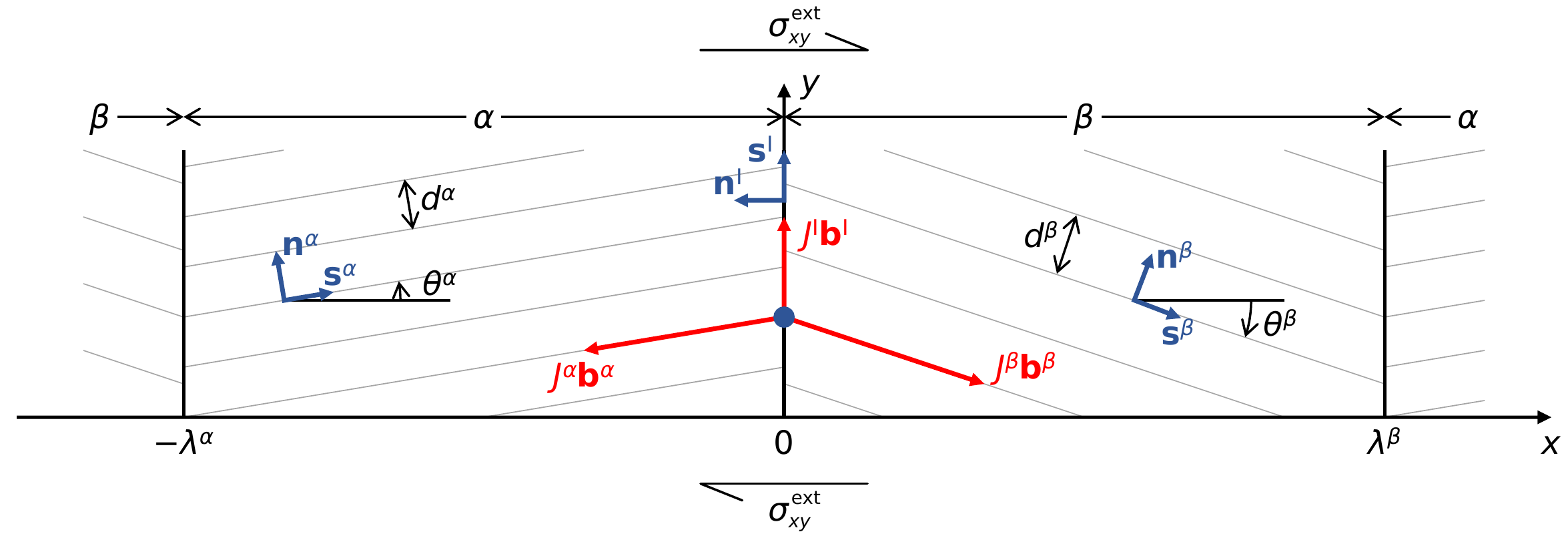}
\caption{\label{fig:bicrystalconfig}
Configuration of a system composed of alternating $A$ and $B$ phases, separated by interfaces, uniform in the $y$-direction, and periodic along  $x$. 
Gray lines denote  slip planes in each phase. 
Blue vectors denote the slip directions or  slip plane normals in each phase. 
The angles $\theta^\alpha$ and $\theta^\beta$ are defined to be positive in a counterclockwise direction.
The  red vectors represent  Burgers vector fluxes at a point on the $\alpha/\beta$ interface (into the two domains and along the interface). 
Fluxes from dislocation reactions (three red vectors) sum to zero.
}
\end{figure}


Consider the simple bicrystal microstructure illustrated in Fig.~\ref{fig:bicrystalconfig}, which represents a 1D bicrystal, periodic along $x$. 
Each period consists of two phases, $\alpha$ and $\beta$ with  domain sizes, $\lambda^{\alpha}$ and $\lambda^{\beta}$. 
Each phase domain is delimited by two, symmetry related  interfaces at $x=0$ and $x=\lambda^{\beta}$. 
For simplicity, assume that there is one slip system in each phase; the slip direction and slip plane normal are $\mathbf{s}^{(i)}$ and $\mathbf{n}^{(i)}$ for $(i) \in\{\alpha,\beta\}$.
To model plastic deformation within the grains, we apply the dislocation-density-function crystal plasticity model of Leung et al.~\citep{leung2015new}.
We denote the  densities of dislocations of opposite sign on each slip system by  subscripts ``$+$'' and ``$-$'' (superscripts denote individual slip systems) such that the dislocation flux 
$\mathbf{J}^{(i)}_{+/-}=\rho^{(i)}_{+/-} \mathbf{v}^{(i)}_{+/-}$, 
where $\mathbf{v}^{(i)}_{+/-}$ is the dislocation velocity. 
In the 1D problem, $\rho^{(i)}$ represents the dislocation density averaged over one period along  $y$ (see \ref{app:onedim}).
We   describe the dislocation velocity magnitude by the power law
\begin{equation}\label{kineticslaw}
v^{(i)}_{+/-}= \pm \mathrm{sgn}(\tau^{(i)}) v^{*(i)} \left|\tau^{(i)}/\tau^{*(i)}\right|^n, 
\end{equation}
where $v^{*(i)}$ and the slip resistance $\tau^{*(i)}$ are  material parameters for phase $(i)$, and 
$n$ is a constant that depends on the range of stress; $n\approx 1$ at low stress~\citep{argon2008strengthening,chang2001dislocation,fan2021strain} and  much higher at large stress~\citep{johnston1959dislocation}.
The velocity law  Eq.~\eqref{kineticslaw}  applies to  grain interiors, while Eq.~\eqref{dissipation2} only describes the dislocation reaction  at the interface. 

The dislocation density evolution satisfies  the balance
\begin{equation}\label{evolution of density}
\dot{\rho}^{(i)}_{+/-}=-\nabla \cdot \mathbf{J}^{(i)}_{+/-}+ \dot{\rho}_{+/-}^{(i),\text{ann}}+ \dot{\rho}_{+/-}^{(i),\text{gen}} 
\end{equation}
(the equation for the 1D problem is in \ref{app:onedim}).
As proposed by Arsenlis et al.~\citep{arsenlis2004evolution}, annihilation of dislocations occurs when two opposite signed dislocations come within a critical capture radius $r_{\uc}$; i.e.,  the annihilation rates are
$\dot{\rho}_{+}^{(i),\text{ann}}
= \dot{\rho}_{-}^{(i),\text{ann}}
= -\rho^{(i)}_{+}\rho^{(i)}_{-}r_{\uc}
\left|v^{(i)}_{+}-v^{(i)}_{-}\right|$.
When a stress is applied, a dislocation pair (opposite signs) is emitted from a source with generation rates~\citep{kocks1975progress}:
$\dot{\rho}_{+}^{(i),\text{gen}}
=\dot{\rho}_{-}^{(i),\text{gen}}
=\eta \left|\tau^{(i)}\right|^{m}$,
where $\eta$ and $m$ are constants. 
Note that in the present model, the sources are assumed to be present everywhere within the grains.
The net dislocation density is 
$\rho^{(i)}=\rho^{(i)}_{+}-\rho^{(i)}_{-}$.
The total RSS $\tau^{(i)}$ in Eq.~\eqref{kineticslaw} has contributions from both the external and internal (associated with all other dislocations) stress tensors ($\boldsymbol{\sigma}^\text{ext}$ and $\boldsymbol{\sigma}^\text{int}$). 
For the simulation results presented here, we apply the external shear stress $\sigma_{xy}^\text{ext}$. 
The internal stress is a functional of dislocation densities on all slip systems (see \ref{app:onedim}).

The flux $J^\txI$ at the interface obeys the reaction law in Eq.~\eqref{dissipation2}, but once dislocations are generated they are assumed to move with velocity $v^\txI$ of the same form as Eq.~\eqref{kineticslaw}, albeit with different parameters:
$v^{\txI}
= \usgn(\tau^{\txI}) v^{*\txI}
\left|\tau^{\txI}/ \tau^{*\txI}\right|^n$,
where we set $n=1$ on all slip systems/interface, although this is not necessary. 
In the simulations, we employ  reduced variables: 
$\tilde{\rho} \equiv \rho (\lambda^{\alpha})^2$,
$\tilde{x} \equiv x/\lambda^{\alpha}$,
$\tilde{t} \equiv v^* t/ \lambda^{\alpha}$, 
$\tilde{v} \equiv v / v^*$, 
$\tilde{\tau} \equiv \tau / K$, 
where $\lambda^{\alpha}$ is the width of the $\alpha$ phase in $x$ (see Fig.~\ref{fig:bicrystalconfig}) and $K \equiv \mu/[2\pi(1-\nu)]$. 
For simplicity, we  omit the tilde in the reduced quantities  below.



\subsection*{Grain Boundaries}

\begin{figure}[h]
\centering
\includegraphics[width=0.99\linewidth]{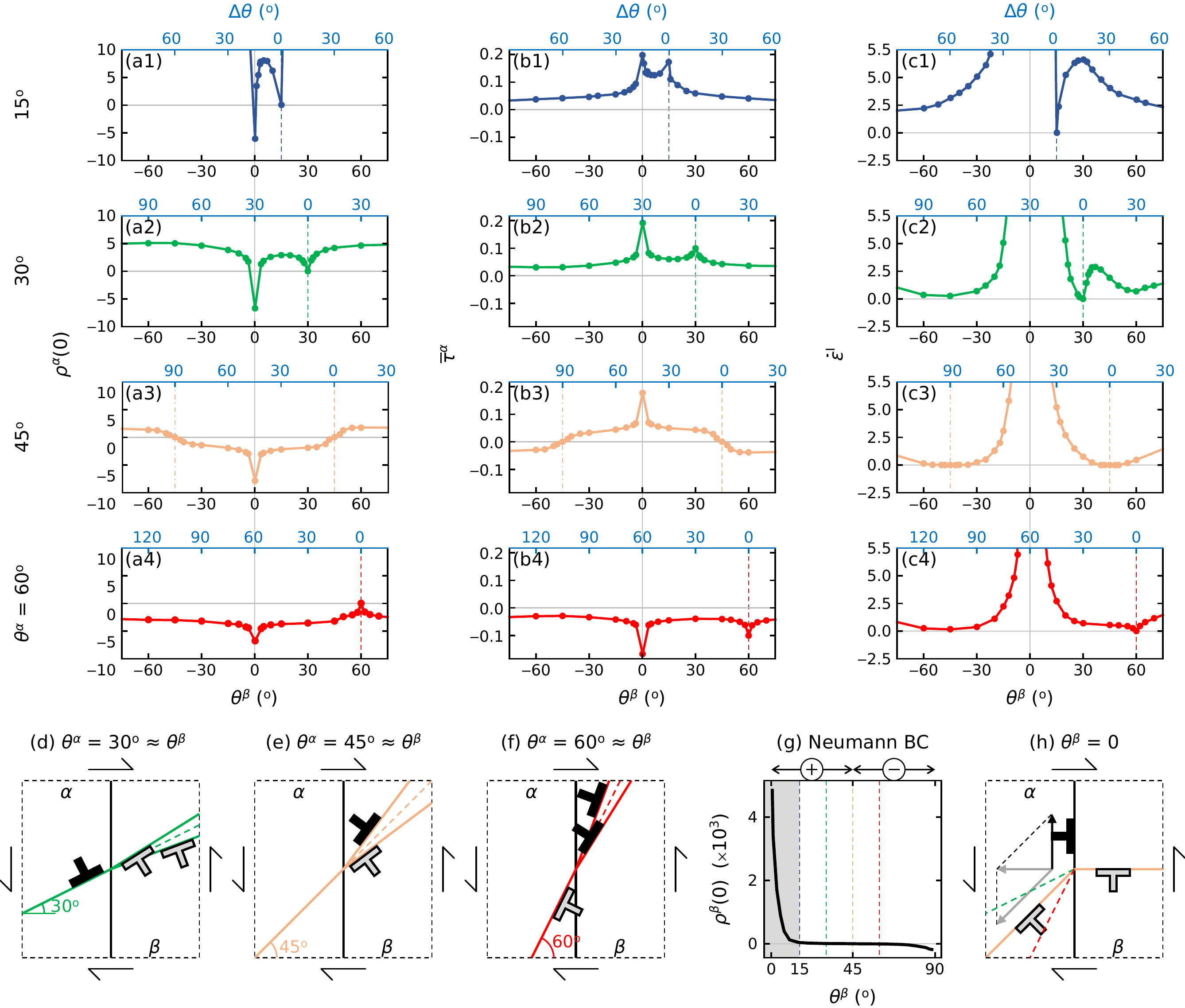}
\caption{\label{fig:dd}
Steady-state bicrystal properties for a range of grain misorientations $\theta^{\alpha}, \theta^{\beta}$ under an external stress $\tau^{\text{ext}}=0.20$.
The four rows represent  fixed $\theta^{\alpha}=15^0,30^0,45^0,60^0$.
(a1)-(a4) Dislocation density $\rho^{\alpha}(0)$ vs. $\theta^{\beta}$.
(b1)-(b4) Total resolved shear stress $\tau^{\alpha}$. 
(c1)-(c4) The strain rate associated with grain boundary sliding  $\dot{\epsilon}^{\txI}=\rho^{\txI} v^{\txI} b^{\txI}=J^{\txI} b^{\txI}$. 
(d)-(f) Schematic of dislocation reactions at the grain boundary for $\theta^\alpha \approx \theta^\beta$.
The dashed and solid lines represent single crystal and interfaces with small misorientations, respectively. 
(g) shows the dislocation pile-up density on the $\beta$ side of the interface $\rho^{\beta}(0)$ vs. the $\beta$ crystal slip system orientation $\theta^{\beta}$ for the no reaction case $\kappa=0$ (Neumann BC).
(h) Schematic of dislocation reactions at the grain boundary when $\theta^\beta$ equals to zero for different $\theta^\alpha$ denoted by different colors. 
}
\end{figure}

When phases $\alpha$ and $\beta$  represent the same structures (but differently oriented), the interface is a grain boundary (GB). 
GBs strongly affect  the mechanical properties of polycrystals~\citep{hall1951deformation,petch1953cleavage}. 
On the other hand, different GBs (different misorientations and inclinations) interact with dislocations  differently~\citep{abuzaid2012slip}, which suggests that the mechanical properties of polycrystals may be adjusted by GB engineering. 
Since  GB properties are functions of misorientation and inclination (and bonding), we examine the effects of $\theta^{\alpha}$ and $\theta^{\beta}$ (Fig.~\ref{fig:bicrystalconfig}) and the interface properties through the coefficient tensor $\mathbf{L}$ in Eq.~\eqref{dissipation2}. 
Upon application of an external shear stress $\sigma_{xy}^{\text{ext}}=0.20$ (see Fig.~\ref{fig:bicrystalconfig}), the dislocation density profile, stress within the bicrystal, and strain evolve.
Figures~\ref{fig:dd}a-c show (i) the dislocation density at the interface on the $\alpha$ side of the interface $\rho^\alpha(x=0)$, (ii) the average resolved shear stress in Phase $\alpha$ $\bar{\tau}^\alpha$, and (iii) the  strain rate associated with GB sliding $\dot{\epsilon}^\txI$ in steady-state (after the transients have relaxed). 
The GB sliding strain rate is $\dot{\epsilon}^\txI\equiv\llbracket{\dot{u}_y(x=0)}\rrbracket/\delta^\txI\equiv \rho^\txI b^\txI v^\txI$, where the term in double brackets is the jump in the $y$-displacement rate across the interface and $\delta^\txI$ is the interface width.

When $\theta^{\beta} = \theta^{\alpha}$, the system is a single crystal.  
This case is indicated by the vertical dashed lines in Figs.~\ref{fig:dd}a-c. 
Not surprisingly,  the net dislocation density at $x=0$ is zero  $\rho^\alpha(0)=0$  (Fig.~\ref{fig:dd}a) and no  sliding $\dot{\epsilon}^\txI = 0$ (Fig.~\ref{fig:dd}c), as expected since there is no GB. 


For small misorientations  $\Delta\theta \equiv |\theta^\alpha - \theta^\beta|$ , the magnitude of the dislocation pileup at the GB  $|\rho^\alpha(0)|$ and  GB sliding rate both increase with increasing misorientation. 
When $\theta^\alpha < 45^\circ$, $\rho^\alpha(0)>0$ (in the region close to the  dashed lines in Fig.~\ref{fig:dd}a1 and a2). 
This may be understood by reference to  schematic Fig.~\ref{fig:dd}d. 
Inside  $\alpha$, the applied stress drives positive dislocations to the GB which react with negative dislocations drawn to the GB from the $\beta$ grain to produce zero pile-up ($\rho(0)=0$) when $\Delta \theta = 0$;   increasing  $\Delta \theta$ increases the positive-dislocation pileup. 
When $\theta^\alpha > 45^\circ$, $\rho^\alpha(0)<0$ (near the red dashed line in Fig.~\ref{fig:dd}a4). 
Figure~\ref{fig:dd}f illustrates that when $\theta^\alpha = 60^\circ$, the applied stress leads to the pileup of negative dislocations; the pileup increases with  increasing $\Delta \theta$. 
When $\theta^\alpha = 45^\circ$, the resolved shear stress is zero and no dislocations are generated in  $\alpha$. 
The dislocation density on the $\alpha$ side of the GB $\rho^\alpha(0)$ is the result of dislocation reactions at the GB. 
As  schematic Fig.~\ref{fig:dd}e shows, when $\theta^\beta > \theta^\alpha$, the positive dislocations pile up on the $\beta$ side of the GB with some transferring into the $\alpha$ phase; this is consistent with the $\theta^\beta > 45^\circ$ results in Fig.~\ref{fig:dd}a3. 
When $\theta^\beta < \theta^\alpha$, some negative dislocations are ``transmitted'' from the $\beta$  to  $\alpha$ phases by reaction, consistent with the data for $\theta^\beta$ smaller than  $45^\circ$ in Fig.~\ref{fig:dd}a3.

We also observe a cusp in $\rho(0)$ at $\theta^\beta = 0$ for all $\theta^\alpha$ (see Fig.~\ref{fig:dd}a). 
When $\theta^\beta = 0$, the resolved shear stress in $\beta$ and the pileup on the $\beta$ side of the GB, $\rho^\beta(0)$, are maximal (since the slip plane is aligned with the external shear stress).  
Figure~\ref{fig:dd}g shows the pileup at the GB for the special  no reaction case $\kappa=0$ (i.e., a Neumann BC). 
For this case, the pileup increases sharply for $\theta < 15^\circ$. 
As   shown schematically in Fig.~\ref{fig:dd}h, when $\theta^\beta=0$, $\rho^\beta(0)$ is very large (corresponding to the sharp peak at $\theta=0$ in Fig.~\ref{fig:dd}g).
Burgers vector reaction then occurs at the GB to produce dislocations (density) that  dominate that on the $\alpha$ side of the GB.
From the Burgers vector reaction shown in Fig.~\ref{fig:dd}h, the dislocation density on the $\alpha$ side of the GB $\rho^\alpha(0)$ is a reaction product and is always negative.

\begin{figure}[t]
\centering
\includegraphics[width=0.95\linewidth]{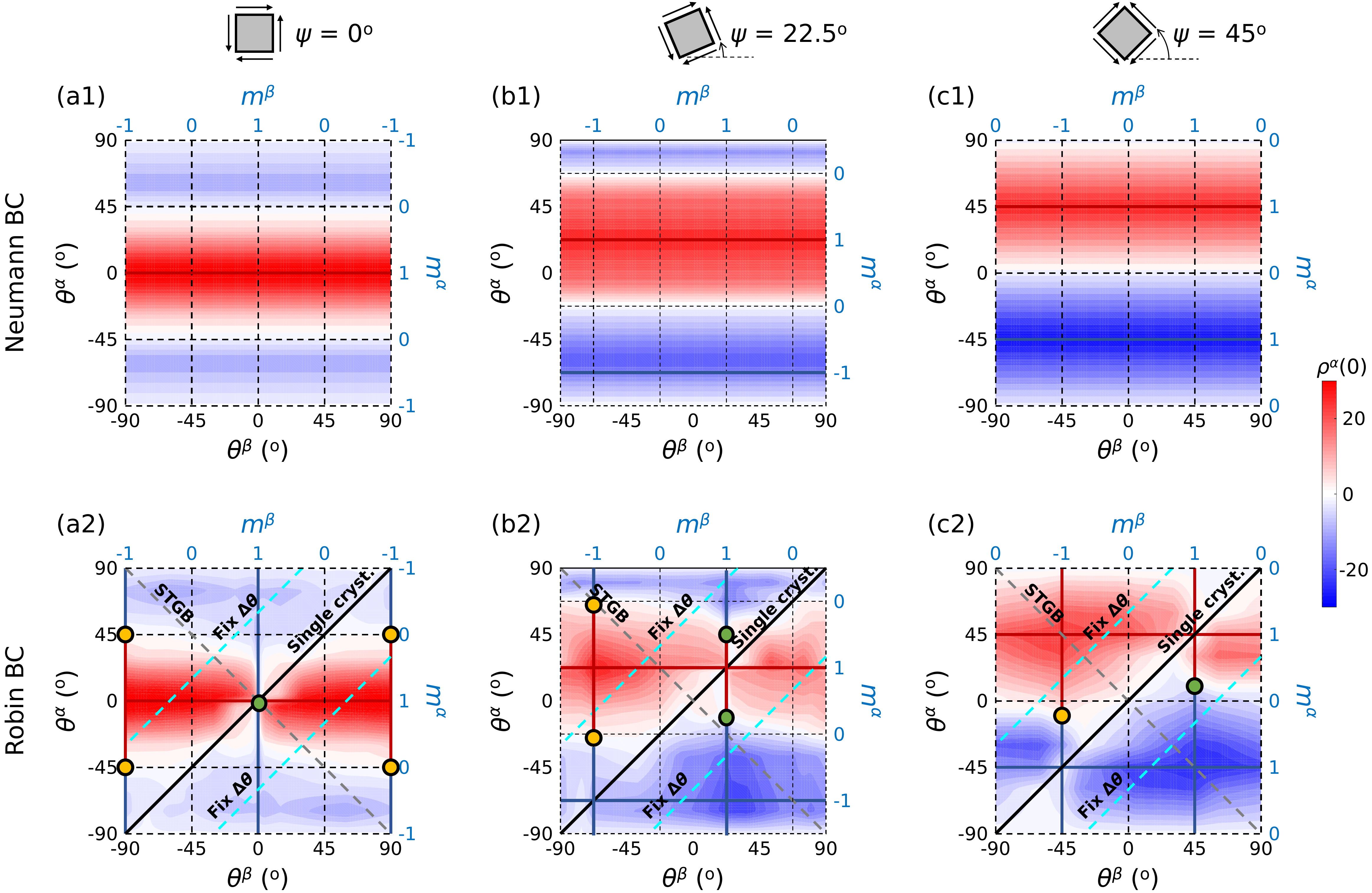}
\caption{\label{fig:evolution of dd}
Interfacial dislocation density $\rho^\alpha(0)$ maps for  shear orientations (a) $\psi=0$, (b) $\psi = 22.5^\circ$ and (c) $\psi = 45^\circ$. 
The top and bottom rows represent  simulation results under no reaction $\kappa=0$ (Neumann) and reaction $\kappa>0$  (Robin) BCs. 
The red/blue color represents positive/negative dislocation pileups at the interfaces.
Maps of $\rho^\beta(0)$ may be  obtained by mirroring the $\rho^\alpha(0)$ maps about $\theta^\alpha = - \theta^\beta$.
Note that the Schmid factors for  $\alpha$ and $\beta$ grains are indicated in blue on the right/vertical  and top/horizontal axes, respectively.
}
\end{figure}

We examine the effects of changing the orientation of the externally applied shear stress (see the top of Fig.~\ref{fig:evolution of dd})  by varying $\psi$ ($0\leq\psi\leq45^{\circ}$) for a set of bicrystals ($\theta^\alpha$, $\theta^\beta$). 
For a GB, $\alpha$ and $\beta$ are the same material, such that we need only   consider one side of the GB (phase $\alpha$ here). 
Figures~\ref{fig:evolution of dd}a-c show the dislocation density $\rho^{\alpha}(0)$ with shear orientation $\psi=0^\circ$, $22.5^\circ$ and $45^\circ$, respectively. 
The two rows of figures represent the results under no-reaction ($\kappa=0$, Neumann) and reaction ($\kappa>0$, Robin -- Eq.~\eqref{dissipation4}) boundary conditions.

No dislocations react at the GB when $\kappa=0$ (Neumann BC).
In this case, the dislocation density $\rho^{\alpha}(0)$ is an extremum when the Schmid factor $m^{\alpha}=\pm1$; and $\rho^{\alpha}(0)$ is zero with $m^{\alpha}=0$ (see Fig.~\ref{fig:evolution of dd}). 
When the reaction BC (Eq.~\eqref{dissipation4}) is applied, the dislocation pileup on the $\beta$ side  affects $\rho^{\alpha}(0)$. 
When $\theta^\alpha=\theta^\beta$, the system is a single crystal/there is no GB (see the solid black lines in Fig.~\ref{fig:evolution of dd}a2-c2), there is no pileup.
Focusing on the $\psi=0$ reaction BC case (Fig.~\ref{fig:evolution of dd}a2) as an example, for fixed $\theta^\alpha$ (any horizontal line), $\rho^{\alpha}(0)$ is an extremum when the Schmid factor  is an extremum, $m^{\beta}=\pm1$. 
When the Schmid factor is an extremum, the magnitudes of the dislocation pileup $\rho^{\beta}(0)$ and $\rho^\alpha(0)$ are extrema.
This is a result of  dislocation reactions at the interface; e.g., at $\theta^{\alpha}=15^{\circ}$ $\rho^{\alpha}(0)$ is nearly zero because of reactions with dislocations from $\beta$ which are very high density at $\theta^{\beta}=0/m^{\beta}=1$.

When the applied shear orientation $\psi$ increases, the $\rho^{\alpha}(0)$ map evolves as seen in  Fig.~\ref{fig:evolution of dd}. 
There are several special points  lying on the lines corresponding to $m^\beta=\pm1$ that represent  transitions between positive and negative values of $\rho^\alpha(0)$.
For the reaction BC cases in Fig.~\ref{fig:evolution of dd}, the special points for $m^\beta = -1$ are indicated by the orange points while those for $m^\beta = 1$ by the green points. 
All of the special points correspond to  $\rho^\alpha(0)=0$.
The special points, along with the $\rho^\alpha(0)=0$ contours, shift towards the upper right within increasing applied shear orientation from $\psi= 0$ to $45^\circ$. 
While dislocation pile-up maps may be generated for any GB and applied shear stress orientation, knowledge of these points and how they change with $\psi$ provide heuristic guidance. 
 
All  symmetric tilt GB (STGB) are located along the diagonal line $\theta^\alpha=-\theta^\beta$ in Fig.~\ref{fig:evolution of dd}.  
For any fixed misorientation $\Delta\theta$, diagonal lines with slope one represent all possible GB inclinations; e.g., see the cyan dashed lines in Figs.~\ref{fig:evolution of dd}a2, b2 and c3. 
The  dislocation density $\rho^\beta(0)$ maps may be  obtained by mirroring  $\rho^\alpha(0)$ maps about $\theta^\alpha=-\theta^\beta$.


\subsection*{Comparison with the LRB Criteria}\label{discussion}

The LRB criteria are widely used to determine the likelihood of slip transfer across a grain boundary~\citep{clark1992criteria,bieler2014grain,han2018grain}. 
These criteria are empirical; they are deduced from extensive experimental observations and simple crystallographic ideas.  
To evaluate slip transfer, we focus on the question of how the dislocation density changes across the GB when transmission occurs.
First, note that even if there are no dislocation reactions at the GB, the plasticity in one grain affects that in the other through the stress concentrations associated with dislocation pile-ups.
Hence, we focus on the change in dislocation density at the interface between cases where slip can occur $\kappa>0$ and cannot occur $\kappa=0$: $\Delta \rho = \rho_{\kappa>0} - \rho_{\kappa=0}$.
Next, we realize that $\Delta \rho$ will be different on the two sides of the interface (depending on grain orientations with respect to one another and the applied stress); hence, $\Delta \rho^{(i)} = \rho^{(i)}_{\kappa>0}- \rho^{(i)}_{\kappa=0}$, where $(i)\in\{\alpha,\beta\}$.
We now compare the LRB predictions with our simulation observations for $\Delta \rho^{(i)}$.

\begin{figure*}[t]
\centering
\includegraphics[width=0.8\linewidth]{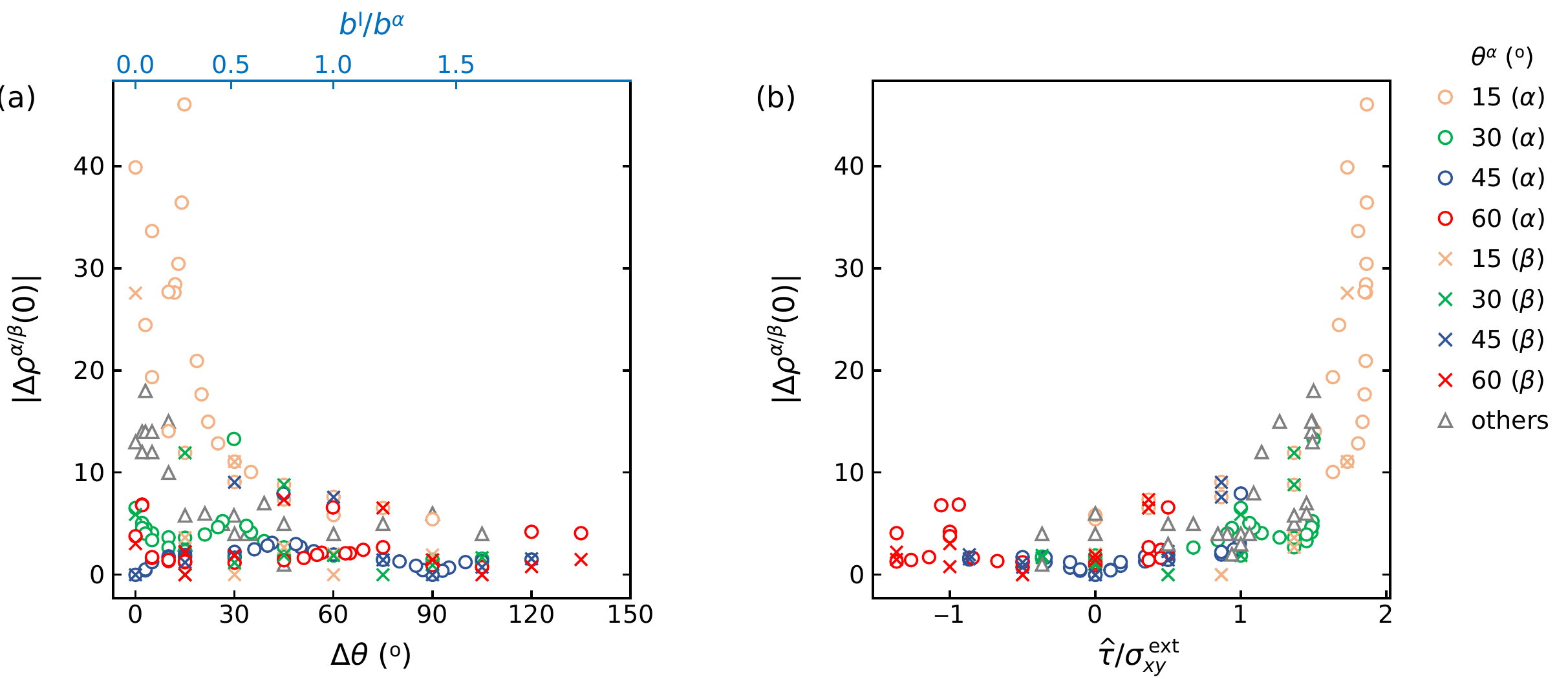}
\caption{\label{fig:residual_burgers}
The change in dislocation density at the GB with the onset of transmission ($\kappa>0$) $|\Delta \rho|$ for several external stress orientations ($\psi=0$, $22.5^\circ$ and $45^\circ$) and grain orientations.
(a) $|\Delta \rho^\text{(i)}|$ versus misorientation  for across the interface $(i)={\alpha}$ (denoted by  ``$\circ$") and $(i)={\beta}$  (denoted by ``$\times$") and residual interfacial Burgers vector magnituted $b^{\txI}/b^{\alpha}$.
(b) The change in dislocation density at the GB $|\Delta \rho^{\alpha}|$ and $|\Delta \rho^{\beta}|$ versus the combined resolved shear stress $\hat{\tau}$ for all different shear orientations.
}
\end{figure*}

The first LRB criterion states that  slip transfer tends to occur between the pair of slip systems with the minimal misorientation angle. 
Our interface BC, Eq.~\eqref{dissipation4}, fully incorporates  the differences in orientation between the two crystals  through the tensor $\mathbf{c} \otimes \mathbf{c}$; hence, the  interface BC may be used to evaluate the first LRB \textit{ansatz}.  
Figure~\ref{fig:residual_burgers}a shows that, in general, $|\Delta \rho^{(i)}|$ decreases with  increasing misorientation angle ($\Delta \theta \equiv |\theta^{\alpha} - \theta^{\beta}|$). 
However, this  trend is \emph{not} universal; we see many examples  in Fig.~\ref{fig:residual_burgers}a where small $\Delta \theta$ corresponds to small $|\Delta\rho^{(i)}|$. 
Of course, when the two slip systems each have small Schmid factors, little slip transfer occurs; this effect is included in the third LRB criterion and shows that the three LRB criteria are not necessarily consistent with each other. 
Therefore, our simulation results based on the interface BC suggest that the first LRB criterion  predicts the correct trends but fails in very many particular cases.


The second LRB criterion states that  slip transfer  occurs in a manner that leads to the smallest residual Burgers vector at the interface. 
The residual Burgers vector associated with any reaction/slip transfer event is simply related to the difference in the Burgers vectors on the two slip systems $\mathbf{b}^\text{I} = \mathbf{b}^\alpha-\mathbf{b}^\beta$, where $\mathbf{b}^\text{I}$ is simply ${b}^\text{I}$ directed along the GB. 
For a GB, this is related to the misorientation angle by $b^\text{I} = 2b^\alpha\sin(\Delta\theta/2)$ if $b^\alpha = b^\beta$ (e.g., for a GB). 
In  Fig.~\ref{fig:residual_burgers}a, we  plot the transmited dislocation density $|\Delta\rho^{(i)}|$ versus $b^\text{I}$ and find that, in general, $|\Delta\rho^{(i)}|$ decreases with  increasing $b^\text{I}$. 
While this trend is consistent with the second LRB criterion, it too fails in many specific cases.  

The third criterion states that the two slip systems involved in slip transfer are those for which the resolved shear stress in the phases/grains is a maximum. 
Abuzaid et al.~\citep{abuzaid2012slip} suggested the use of the combined resolved shear stresses in the two phases/grains for this condition:
\begin{equation}\label{Schmid factor1}
\hat{\tau} 
\equiv (\mathbf{s}^{\alpha} \otimes \mathbf{n}^{\alpha} + \mathbf{s}^{\beta} \otimes \mathbf{n}^{\beta}) : \boldsymbol{\sigma}=\tau^\alpha+\tau^\beta.
\end{equation}
Figure~\ref{fig:residual_burgers}b shows  $|\Delta\rho^{(i)}|$ versus $\hat{\tau}$ from the simulations. 
We observe that $|\Delta\rho^{(i)}|$  indeed increases with  increasing $\hat{\tau}$. 
This  (third)  criterion  properly describe the trends, but, again, does \emph{not} always work. 
Based on the interface BC (Eq.~\eqref{dissipation4}), when the interface dislocation density is small ($\rho^{\txI}\rightarrow0$), the flux on the $\alpha$ side of the GB is
\begin{equation}\label{Schmid factor2}
J^{\alpha} = \kappa (c^{\beta \txI})^2 \rho^{\alpha} \tau^{\alpha} + \kappa (c^{\beta \txI} c^{\txI \alpha}) \rho^{\beta} \tau^{\beta}.
\end{equation}
For the very special case of a symmetric tilt grain boundary, $c^{\txI \alpha}=c^{\beta \txI}$ and $\rho^{\alpha}=\rho^{\beta}$ for which Eq.~\eqref{Schmid factor2} reduces to 
\begin{equation}\label{LRB3}
J^{\alpha} = \kappa (c^{\beta \txI})^2 \rho^{\alpha} \hat{\tau}.
\end{equation}
This shows that, for this special case (STGB), the maximum combined resolved shear stress $\hat{\tau}$ produces the maximum flux and the third LRB criterion is exactly correct, while our interface BC is applicable in general (symmetric and asymmetric GBs; heterophase interfaces).  

The three LRB criteria are insightful, but largely heuristic and are often not  consistent with one another.  
For example, consider the case of applying an external stress $\sigma_{xy}^\text{ext}$ ($\sigma_{xx}^\text{ext}=\sigma_{yy}^\text{ext}=0$), keeping $\theta^{\alpha}=0$, and gradually increasing $\theta^{\beta}$ from 0 to $90^\circ$. 
Since the combined resolved shear stress is a minimum at $\theta^{\beta} \approx 45^\circ$, the third LRB criterion suggests that the transmitted dislocation density $|\Delta\rho^{(i)}|$ should also be a minimum at this angle. 
On the other hand, when $\theta^{\beta}$ gradually increases from 0 to $90^\circ$, the misorientation angle between the two slip systems $\Delta\theta$ increases monotonically and, according to the first criterion, $|\Delta\rho^{(i)}|$ will decrease monotonically. 
This means that $|\Delta\rho^{(i)}|$ does not reach a minimum at $\theta^{\beta} \approx 45^\circ$. 
This simple example demonstrates that the LRB criteria are not self-consistent, unlike our interface BC.

\section{Conclusion}
We proposed a meso-scale  boundary condition to model the interactions between dislocations from within grains and interfaces. 
Our interface BC, based on interface bicrystallography and rigorous linear kinetics, is  novel, self-consistent, and easily applied. 
Consider  the following: 
\begin{itemize}[noitemsep,nolistsep]
\item[(i)] 
The interface BC is established based upon experimentally observed reactions between  lattice and interface dislocations/disconnections. 
Burgers vectors are rigorously conserved. 
\item[(ii)]
The interface BC is based on basic kinetic theory (the principle of maximum dissipation rate) for dislocation-interface interactions (rather than energy minimization, as in some other models). 
\item[(iii)]
The interface BC is applicable to interfaces in all crystalline systems  and  to multiple slip systems. 
\item[(iv)]
Interface sliding  naturally occurs in simulations incorporating the interface BC (i.e., interface dislocations participate in reactions at the interface). 
\end{itemize}

We employed the proposed interface BC to examine the validity of the empirical LRB criteria for slip transfer across a GB. 
We demonstrated that the three LRB criteria correctly  predict the  slip transfer trends but the LRB criteria fail in many cases and are not self-consistent. 
The interface BC provides a more rigorous and accurate approach to consider all factors that affect slip transfer, including  all bicrystallography (including misorientation angle), the residual Burgers vectors, interface sliding and arbitrary external  stress. 
The proposed interface BC can be applied directly in most plasticity  simulation methods, such as continuum dislocation dynamics, discrete dislocation dynamics and crystal plasticity finite element method. 

\section*{Acknowledgements}
JY, DJS and JH were also supported by the  National Key R\&D Program of China (2021YFA1200202). 
JY, AHWN and DJS  gratefully acknowledge support of the Hong Kong Research Grants Council Collaborative Research Fund C1005-19G. 
JH acknowledges support of Early Career Scheme (ECS) Grant from the Hong Kong Research Grants Council 21213921. 
AHWN also acknowledges support from the Shenzhen Fund 2021 Basic Research General Programme JCY20210324115400002 and the  Guangdong Province Basic and Applied Research Key Project 2020190718102.

\appendix
\section{Linear response theory}
\label{linearresponse}

The linear response approach, underlying Eq.~\eqref{dissipation2}, may be deduced based on the maximum entropy production principle~\citep{ziegler2012introduction,martyushev2006maximum} or equivalently the principle of maximum dissipation rate~\citep{onsager1931reciprocal}. 
The general idea behind the  derivation is as follows. 
First, the global entropy production rate is a functional of the (generalized) flux: $\dot{\Sigma} = \dot{\Sigma}[\mathbf{J}]$. 
Second, we assume that local thermodynamic equilibrium ($1^\text{st}$ law of thermodynamics) applies through the constraint: $\Gamma[\mathbf{J}] = 0$. 
Finally, we maximize the global entropy production rate $\dot{\Sigma}$ with respect to $\mathbf{J}$ under the  $\Gamma=0$ constraint via the Lagrange multiplier method: $\delta(\dot{\Sigma} - \lambda\Gamma) = 0$, where $\lambda$ is the Lagrange multiplier. 
The three steps are detailed below. 

Since any nonequilibrium process is characterized by the presence of a flux, the local entropy production rate is a function of the flux: $\dot{\sigma} = \dot{\sigma}(\mathbf{J})$, where $\mathbf{J}$ is a vector of the fluxes along all slip systems into the interface. 
When the system is near equilibrium, we  expand $\dot{\sigma}(\mathbf{J})$ about $\mathbf{J} = \mathbf{0}$. 
From  symmetry considerations, $\dot{\sigma} = \mathbf{J}\cdot\mathbf{C}\mathbf{J}$, where $\mathbf{C}$ is a coefficient tensor. 
The global entropy production rate is 
\begin{equation}\label{SigmaJintJCJdV}
\dot{\Sigma}[\mathbf{J}] = \int \mathbf{J}\cdot\mathbf{C}\mathbf{J} \rmd V. 
\end{equation}

Local thermodynamic equilibrium implies we may write the entropy production rate as
\begin{equation}\label{localequilibrium}
\ud u = T\ud s + \sum_i \phi^{(i)} \ud \rho^{(i)}
\quad\Rightarrow\quad
\dot{s} = \frac{1}{T}\dot{u} + \sum_i \left(-\frac{\phi^{(i)}}{T}\right) \dot{\rho}^{(i)},
\end{equation}
where $u$ is the internal energy density, $T$ is the temperature, $s$ is the entropy density; 
$\phi^{(i)} = \phi^{(i)}(\zeta^{(i)})$ is the free energy of a dislocation (per length) on the $i^\text{th}$ slip system and located at $\zeta^{(i)}$, where $\zeta^{(i)}$ is the coordinate along the $\boldsymbol{\zeta}^{(i)}$-axis (see Fig.~\ref{fig:bc_3d}) and 
$\rho^{(i)} = \rho^{(i)}(\zeta^{(i)})$ is the dislocation density at $\zeta^{(i)}$. 
The continuity equations at each point on the interface are  
\begin{equation}\label{continuityeq}
\dot{s} = \dot{\sigma} - \sum_i \frac{\partial J_s^{(i)}}{\partial \zeta^{(i)}}, 
\quad
\dot{u} = - \sum_i \frac{\partial J_u^{(i)}}{\partial \zeta^{(i)}}, 
\quad
\dot{\rho}^{(i)} = - \frac{\partial J^{(i)}}{\partial \zeta^{(i)}},
\end{equation}
where $J_s^{(i)}$, $J_u^{(i)}$ and $J^{(i)}$ are, respectively, the entropy flux , the energy flux and the dislocation flux flowing from the $i^\text{th}$ slip system to the interface point. 
Substituting Eq.~\eqref{continuityeq} into Eq.~\eqref{localequilibrium}, we find 
\begin{equation}\label{dotsigmasumiJui}
\dot{\sigma}
=
\sum_i \left[
J_u^{(i)} \frac{\partial}{\partial\zeta^{(i)}}
\left(\frac{1}{T}\right)
+
J^{(i)} \frac{\partial}{\partial\zeta^{(i)}}
\left(-\frac{\phi^{(i)}}{T}\right)
\right].  
\end{equation}
We define the generalized flux as $\mathbf{J} \equiv (J_u^{(1)}, \cdots, J_u^{(4)}, J^{(1)}, \cdots, J^{(4)})^\text{T}$ and the generalized force as 
\begin{equation}
\mathbf{f} 
\equiv \left(
\frac{\partial}{\partial\zeta^{(1)}}\left(\frac{1}{T}\right), 
\cdots, 
\frac{\partial}{\partial\zeta^{(4)}}\left(\frac{1}{T}\right), 
\frac{\partial}{\partial\zeta^{(1)}}\left(-\frac{\phi^{(1)}}{T}\right), 
\cdots,
\frac{\partial}{\partial\zeta^{(4)}}\left(-\frac{\phi^{(4)}}{T}\right)
\right)^\text{T}.
\end{equation}
Then, Eq.~\eqref{dotsigmasumiJui} can be written as $\dot{\sigma} = \mathbf{J}\cdot\mathbf{f}$. 
Thus,  local thermodynamic equilibrium leads to the constraint: 
\begin{equation}\label{GammaJintfJJCJdV}
\Gamma[\mathbf{J}] = \int (\mathbf{J}\cdot\mathbf{f} - \mathbf{J}\cdot\mathbf{C}\mathbf{J})\rmd V = 0. 
\end{equation}

To maximize Eq.~\eqref{SigmaJintJCJdV} under the constraint Eq.~\eqref{GammaJintfJJCJdV}, we construct the functional:
\begin{equation}
\dot{\Sigma}'[\mathbf{J}]
= \int \mathbf{J}\cdot\mathbf{C}\mathbf{J} \rmd V
+ \int \lambda(\mathbf{J}\cdot\mathbf{f} - \mathbf{J}\cdot\mathbf{C}\mathbf{J})\rmd V, 
\end{equation}
and set $\delta\dot{\Sigma}'/\delta\lambda = 0$ and $\delta\dot{\Sigma}'/\delta\mathbf{J} = \mathbf{0}$. 
The solution to this variational problem is $\mathbf{J} = \mathbf{L}\mathbf{f}$, where $\mathbf{L}\equiv \mathbf{C}^{-1}$; thus, we have obtained the linear response expression, Eq.~\eqref{dissipation2}. 
Note that the assumptions include: (i) near-equilibrium and (ii) local equilibrium. 
It remains to examine the physical meaning of the force $\mathbf{f}$. 
Recall that $f^{(i)} \equiv -(1/T)(\partial\phi^{(i)}/\partial\zeta^{(i)})$. 
The derivative $-\partial\phi^{(i)}/\partial\zeta^{(i)}$ represents the decrease of the free energy of a dislocation (per length) when the dislocation is displaced by a small distance $\rmd \zeta^{(i)}$. 
So, $f^{(i)}T$ exactly corresponds to the Peach-Koehler force. 
In Eq.~\eqref{dissipation2}, $T$ is absorbed into the coefficient tensor such that the $\{f^{(i)}\}$ represent the Peach-Koehler force.

\section{Derivation of the interface boundary condition for multiple slip systems}\label{multi}

Extension of the interface BC to the case of multiple slip systems in both phases is given below. 
Consider the case where there are multiple non-colinear slip systems in both phases and one slip plane along the interface. 
As an example, assume that there are two slip systems in the $\alpha$ phase, ``$\alpha_1$'' and ``$\alpha_2$'', and two in  $\beta$, ``$\beta_1$'' and ``$\beta_2$''. 
Including the interface ``I'', there are five slip systems. 
Label the five slip systems as 
$\text{``1''} \equiv \text{``$\alpha_1$''}$, 
$\text{``2''} \equiv \text{``$\alpha_2$''}$, 
$\text{``3''} \equiv \text{``$\beta_1$''}$, 
$\text{``4''} \equiv \text{``$\beta_2$''}$, and 
$\text{``5''} \equiv \text{``I''}$. 
Reactions may occur amongst any four of the five slip systems. 
For example, 
\begin{equation}\label{J1245}
J^{(1)} b^{(1)} \mathbf{s}^{(1)} + J^{(2)} b^{(2)} \mathbf{s}^{(2)} 
+ J^{(4)} b^{(4)} \mathbf{s}^{(4)}
+ J^{(5)} b^{(5)} \mathbf{s}^{(5)} 
= \mathbf{0}.
\end{equation}
Below, we enumerate all possible reactions:
\begin{equation}
\begin{array}{lll}
\text{Reaction 1:}~~(1)+(2)+(3)+(4)=0, \qquad
\text{Reaction 2:}~~(1)+(2)+(3)+(5)=0, &\\
\text{Reaction 3:}~~(1)+(2)+(4)+(5)=0, \qquad
\text{Reaction 4:}~~(1)+(3)+(4)+(5)=0, &\\
\text{Reaction 5:}~~(2)+(3)+(4)+(5)=0,\qquad &
\end{array}
\nonumber
\end{equation}
where ``$(i)$'' is short for ``$J^{(i)}b^{(i)}\mathbf{s}^{(i)}$'' for the $i^\text{th}$ slip system. 
Take Eq.~\eqref{J1245} (Reaction 3) as an example. 
Similar to Eq.~\eqref{dissipation4}, Reaction 3 kinetics may be described by
\begin{equation}\label{J9BPCBTr0}
\left(\begin{array}{c}
J^{(1)}_3b^{(1)} \\
J^{(2)}_3b^{(2)} \\ 
J^{(4)}_3b^{(4)} \\ 
J^{(5)}_3b^{(5)}
\end{array}\right)
=
\left(\begin{array}{cccc}
L_{11} & L_{12} & L_{14} & L_{15}\\
 & L_{22} & L_{24} & L_{25}\\
 & & L_{44} & L_{45}\\
\text{sym.} & & & L_{55} 
\end{array}\right)
\left(\begin{array}{c}
\tau^{(1)} \rho^{(1)} b^{(1)} \\
\tau^{(2)} \rho^{(2)} b^{(2)}\\ 
\tau^{(4)} \rho^{(4)} b^{(4)}\\ 
\tau^{(5)} \rho^{(5)} b^{(5)}
\end{array}\right),
\end{equation}
where $J^{(i)}_l$ is the flux on slip system $i$ generated by Reaction $l$.
We can rewrite this equation as
\begin{align}\label{J9BPCBTr}
&\left(\begin{array}{c}
J^{(1)}_3b^{(1)} \\ 
J^{(2)}_3b^{(2)} \\ 
J^{(3)}_3b^{(3)} \\
J^{(4)}_3b^{(4)} \\ 
J^{(5)}_3b^{(5)} \\ 
\end{array}\right)
=
\kappa_3
\left(\begin{array}{ccccc}
(c^{(245)})^2 & c^{(245)}c^{(415)} & 0 & c^{(245)}c^{(125)} & c^{(245)}c^{(142)} \\
 & (c^{(415)})^2 & 0 & c^{(415)}c^{(125)} & c^{(415)}c^{(142)} \\
  &  & 0 & 0 & 0 \\
 & & & (c^{(125)})^2 & c^{(125)}c^{(142)} \\
\text{sym.} & & & & (c^{(142)})^2
\end{array}\right)
\left(\begin{array}{c}
\tau^{(1)} \rho^{(1)} b^{(1)} \\
\tau^{(2)} \rho^{(2)} b^{(2)}\\ 
\tau^{(3)} \rho^{(3)} b^{(3)}\\ 
\tau^{(4)} \rho^{(4)} b^{(4)}\\ 
\tau^{(5)} \rho^{(5)} b^{(5)}
\end{array}\right)
\nonumber\\
\Rightarrow~&
\bar{\mathbf{J}}_{3,\txI}
= \bar{\mathbf{B}}^{-1}\kappa_3\overline{(\mathbf{c}\otimes\mathbf{c})_3} \bar{\mathbf{B}} \bar{\mathbf{T}}_\txI \bar{\boldsymbol{\rho}}_\txI,
\end{align}
where the quantities with overlines are extended to $5\times 1$ vectors or $5\times 5$ matrices with zero padding. 
The tensor $\overline{(\mathbf{c}\otimes\mathbf{c})_3}$ depends on the geometry, which slip systems are available, and which four slip systems participate in Reaction $3$. 
We  obtain the relationship akin to Eq.~\eqref{J9BPCBTr} for all the other reactions. 
The total flux associated with these 5 reactions is 
\begin{equation}\label{apped_5}
\bar{\mathbf{J}}_\txI
= \sum_{l=1}^5 \bar{\mathbf{J}}_{l,\txI}
= \bar{\mathbf{B}}^{-1}
\left[
\sum_{l=1}^5 \kappa_l
\overline{(\mathbf{c}\otimes\mathbf{c})_l}
\right]
\bar{\mathbf{B}}
\bar{\mathbf{T}}_\txI
\bar{\boldsymbol{\rho}}_\txI. 
\end{equation}
This  example is for the five slip systems and five reactions case. 
If there are $N$ slip systems (including the slip systems in the interface) and $M \equiv C_N^4$ reactions, the interface BC becomes Eq.~\eqref{rbc10}.

\section{One-dimensional problem as coarse-graining of the two-dimensional problem}
\label{app:onedim}

\begin{figure}[t]
\centering
\includegraphics[width=0.95\linewidth]{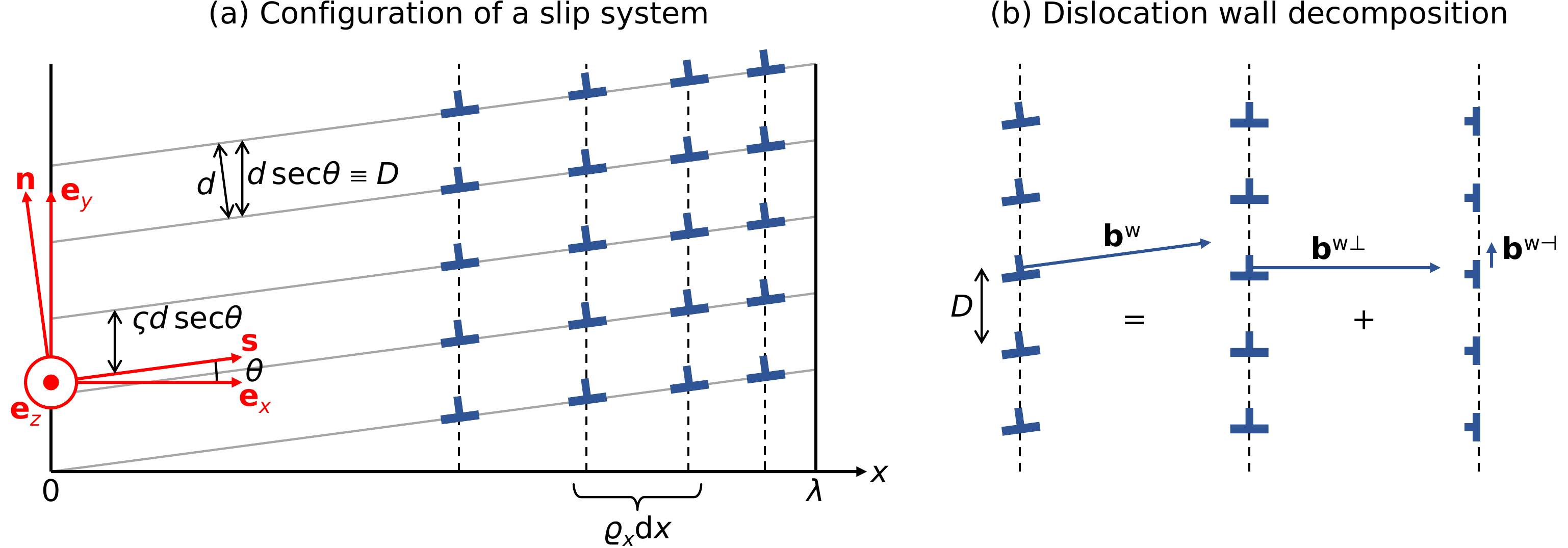}
\caption{\label{fig:diswall}
(a) Geometry of a configuration of parallel slip planes and the coordinate system. 
(b) A dislocation wall may be viewed as composed of two dislocation walls whose stress fields are known. 
}
\end{figure}

A 1D problem can be deduced from a 2D problem by coarse-graining when the dislocation distribution is periodic along the $y$-axis in Fig.~\ref{fig:bicrystalconfig}. 
We focus on the configuration illustrated in Fig.~\ref{fig:diswall}a (i.e., a set of dislocation walls distributed along the $x$-axis). 
Each dislocation wall consists of a periodic array of edge dislocations and each dislocation sits on its own slip plane. 
The dislocation density can be written as 
\begin{equation}\label{rhox}
\rho(x, y)
= \varrho(x) \sum_{j=-\infty}^\infty \delta\left(y - y_j(x)\right), 
\end{equation}
where $\varrho(x)$ is the number density of the dislocation walls distributed along the $x$-axis (i.e., the vertical dashed lines in Fig.~\ref{fig:diswall}a), $y_j(x) = x\tan\theta - (j+\varsigma)d\sec\theta$, $d$ is the interplanar spacing, and $\varsigma \in [0,1)$ is the fractional offset along the $y$-axis at $x=0$. 
We  obtain a 1D quantity by averaging (coarse-graining) the corresponding 2D quantity over a period alone $y$; i.e., $D^{-1}\int_0^D \rmd y$, where $D \equiv d\sec\theta$ is the period along  $y$. 
Thus, we  define the coarse-grained 1D dislocation density as
\begin{equation}
\bar{\rho}
\equiv \frac{1}{D}\int_0^D \rho \rmd y
= \frac{1}{D}\int_0^D \varrho\sum_j\delta(y-y_j) \rmd y
= \frac{\varrho}{D}
= \frac{\varrho\cos\theta}{d}.
\end{equation}
Substituting Eq.~\eqref{rhox} into Eq.~\eqref{evolution of density} and averaging on both sides of Eq.~\eqref{evolution of density} (without the annihilation and generation rates), we have  
\begin{equation}\label{dotvarrhobarv}
\dot{\bar{\rho}}
= - \frac{\partial(\bar{\rho} v_x)}{\partial x} 
= - \cos\theta \frac{\partial(\bar{\rho} v)}{\partial x}. 
\end{equation}
This equation describes the evolution of  dislocation density in 1D. 
(Note that for the 1D problem studied in the main text, we omit the bar of $\bar{\rho}$  for simplicity.) 

Equation~\eqref{dotvarrhobarv} requires the evaluation of $v = \usgn(\tau) v^* |\tau/\tau^*|^n$ and the RSS $\tau = \mathbf{s}\cdot \boldsymbol{\sigma}\mathbf{n}$. 
The total stress $\boldsymbol{\sigma}$ has contribution from the external stress $\boldsymbol{\sigma}^\text{ext}$ and the internal one $\boldsymbol{\sigma}^\text{int}$ associated with the elastic interaction between dislocations. 
$\boldsymbol{\sigma}^\text{ext}$ is a constant in a stress-controlled experiment. 
The problem is how to calculate $\boldsymbol{\sigma}^\text{int}$. 
The internal stress field due to a distribution of dislocations in a 2D space is
\begin{equation}\label{sigmaint0}
\boldsymbol{\sigma}^\text{int}(\mathbf{r})
= \iint \rho(\mathbf{r}') 
\boldsymbol{\sigma}^\txd(\mathbf{r}-\mathbf{r}') \rmd \mathbf{r}',
\end{equation}
where $\boldsymbol{\sigma}^\txd(\mathbf{r}-\mathbf{r}')$ is the stress at the point $\mathbf{r}$ induced by a dislocation located at $\mathbf{r}'$. 
Substituting Eq.~\eqref{rhox} (with $\varsigma=0$) into Eq.~\eqref{sigmaint0}, 
\begin{align}
\boldsymbol{\sigma}^\text{int}(x, y)
&= \iint \varrho(x')
\sum_j \delta(y' - y_j(x'))
\boldsymbol{\sigma}^\txd(x-x', y-y') \rmd x' \rmd y'
\nonumber\\
&= \int \varrho(x')
\sum_j 
\boldsymbol{\sigma}^\txd(x-x', y-x'\tan\theta - jD) \rmd x' 
= \int \varrho(x')
\boldsymbol{\sigma}^\txw(x-x', y-x'\tan\theta; D) \rmd x',  
\end{align}
where $\boldsymbol{\sigma}^\txw(x-x', y-y'; D) \equiv \sum_j\boldsymbol{\sigma}^\txd(x-x', y-y'-jD)$ is the stress field at $(x,y)$ due to a vertical dislocation wall for which the dislocation spacing is $D$ and one of the dislocations is located at $(x', y')$. 
A dislocation wall may be viewed as the superposition of two dislocation walls for which analytical solutions are known; such composition is shown in Fig.~\ref{fig:diswall}b and can be expressed as
\begin{equation}
\boldsymbol{\sigma}^\text{w}(x, y;D)
= 
\boldsymbol{\sigma}^{\txw\perp}(x,y;D)
+ \boldsymbol{\sigma}^{\txw\dashv}(x,y;D),  
\end{equation}
where $\boldsymbol{\sigma}^{\txw\perp/\dashv}$ is resulted from the dislocation wall associated with the Burgers vector $\mathbf{b}^{\txw\perp/\dashv}$. 
The analytical solutions to $\boldsymbol{\sigma}^{\txw\perp}$ and $\boldsymbol{\sigma}^{\txw\dashv}$ are~\citep{anderson2017theory}
\begin{equation}
\left\{
\renewcommand*{\arraystretch}{1.2}
\begin{array}{lll}
\sigma_{xx}^{\txw\perp}
= -\sigma_0 \sin Y \left(
\cosh X - \cos Y + X\sinh X \right)
\\
\sigma_{yy}^{\txw\perp}
= -\sigma_0 \sin Y\left(
\cosh X - \cos Y - X\sinh X \right)
\\
\sigma_{xy}^{\txw\perp}
= \sigma_0 X \left(\cosh X \cos Y - 1\right)
\end{array}\right.
\end{equation}
and
\begin{equation}
\left\{
\renewcommand*{\arraystretch}{1.2}
\begin{array}{lll}
\sigma_{xx}^{\txw\dashv}
= \sigma_0 X \left(
\cosh X \cos Y - 1\right)
\\
\sigma_{yy}^{\txw\dashv}
= \sigma_0 \left[
2\sinh X\left(\cosh X - \cos Y\right) - 
X \left(\cosh X \cos Y - 1\right)
\right]
\\
\sigma_{xy}^{\txw\dashv}
= -\sigma_0 \sin Y\left(
\cosh X - \cos Y - X\sinh X\right)
\end{array}\right.,
\end{equation}
where
\begin{equation}
\sigma_0
\equiv
\frac{\mu b/D}{2(1-\nu)(\cosh X - \cos Y)^2},
\quad
X \equiv \frac{2\pi x}{D},
\quad
Y \equiv \frac{2\pi y}{D}. 
\end{equation}
The resolved internal shear stress is $\tau^\text{int} = \mathbf{s} \cdot  \boldsymbol{\sigma}^\text{int}\mathbf{n}$. 
From this, we write 
\begin{align}
&\tau^\text{int}(x,y)
= 
\int \varrho(x') 
\tau^\text{w}(x-x', y-x'\tan\theta; d\sec\theta) \rmd x',
\quad
\tau^\text{w} = \tau^{\txw\perp} + \tau^{\txw\dashv},
\nonumber\\
&\tau^{\txw\perp}(x,y)
= \sigma_0 X[\sinh X \sin Y \sin 2\theta + (\cosh X \cos Y - 1)\cos 2\theta],
\nonumber\\
&\tau^{\txw\dashv}(x,y)
= \sigma_0 \big\{
[\sinh X (\cosh X - \cos Y) - X(\cosh X \cos Y - 1)]\sin 2\theta
\nonumber\\
&- \sin Y(\cosh X - \cos Y - X\sinh X) \cos 2\theta
\big\}. 
\end{align}
Since all  dislocations sit on the slip planes, we only  evaluate the stress at $(x, x\tan\theta)$; $\tau^\text{int}(x) \equiv \tau^\text{int}(x, x\tan\theta)$.

\bibliographystyle{elsarticle-harv}
\bibliography{mybib}

\providecommand{\noopsort}[1]{}\providecommand{\singleletter}[1]{#1}%
\begin{thebibliography}{77}
\expandafter\ifx\csname natexlab\endcsname\relax\def\natexlab#1{#1}\fi
\providecommand{\url}[1]{\texttt{#1}}
\providecommand{\href}[2]{#2}
\providecommand{\path}[1]{#1}
\providecommand{\DOIprefix}{doi:}
\providecommand{\ArXivprefix}{arXiv:}
\providecommand{\URLprefix}{URL: }
\providecommand{\Pubmedprefix}{pmid:}
\providecommand{\doi}[1]{\href{http://dx.doi.org/#1}{\path{#1}}}
\providecommand{\Pubmed}[1]{\href{pmid:#1}{\path{#1}}}
\providecommand{\bibinfo}[2]{#2}
\ifx\xfnm\relax \def\xfnm[#1]{\unskip,\space#1}\fi
\bibitem[{Abuzaid et~al.(2012)Abuzaid, Sangid, Carroll, Sehitoglu and
  Lambros}]{abuzaid2012slip}
\bibinfo{author}{Abuzaid, W.Z.}, \bibinfo{author}{Sangid, M.D.},
  \bibinfo{author}{Carroll, J.D.}, \bibinfo{author}{Sehitoglu, H.},
  \bibinfo{author}{Lambros, J.}, \bibinfo{year}{2012}.
\newblock \bibinfo{title}{{Slip transfer and plastic strain accumulation across
  grain boundaries in Hastelloy X}}.
\newblock \bibinfo{journal}{Journal of the Mechanics and Physics of Solids}
  \bibinfo{volume}{60}, \bibinfo{pages}{1201--1220}.
\newblock \DOIprefix\doi{https://doi.org/10.1016/j.jmps.2012.02.001}.
\bibitem[{Acharya(2001)}]{acharya2001model}
\bibinfo{author}{Acharya, A.}, \bibinfo{year}{2001}.
\newblock \bibinfo{title}{A model of crystal plasticity based on the theory of
  continuously distributed dislocations}.
\newblock \bibinfo{journal}{Journal of the Mechanics and Physics of Solids}
  \bibinfo{volume}{49}, \bibinfo{pages}{761--784}.
\newblock \DOIprefix\doi{https://doi.org/10.1016/S0022-5096(00)00060-0}.
\bibitem[{Anderson et~al.(2017)Anderson, Hirth and Lothe}]{anderson2017theory}
\bibinfo{author}{Anderson, P.M.}, \bibinfo{author}{Hirth, J.P.},
  \bibinfo{author}{Lothe, J.}, \bibinfo{year}{2017}.
\newblock \bibinfo{title}{Theory of dislocations}.
\newblock \bibinfo{publisher}{Cambridge University Press}.
\bibitem[{Argon(2008)}]{argon2008strengthening}
\bibinfo{author}{Argon, A.S.}, \bibinfo{year}{2008}.
\newblock \bibinfo{title}{Strengthening mechanisms in crystal plasticity}.
  volume~\bibinfo{volume}{4}.
\newblock \bibinfo{publisher}{Oxford University Press on Demand}.
\bibitem[{Arsenlis et~al.(2004)Arsenlis, Parks, Becker and
  Bulatov}]{arsenlis2004evolution}
\bibinfo{author}{Arsenlis, A.}, \bibinfo{author}{Parks, D.M.},
  \bibinfo{author}{Becker, R.}, \bibinfo{author}{Bulatov, V.V.},
  \bibinfo{year}{2004}.
\newblock \bibinfo{title}{On the evolution of crystallographic dislocation
  density in non-homogeneously deforming crystals}.
\newblock \bibinfo{journal}{Journal of the Mechanics and Physics of Solids}
  \bibinfo{volume}{52}, \bibinfo{pages}{1213--1246}.
\newblock \DOIprefix\doi{https://doi.org/10.1016/j.jmps.2003.12.007}.
\bibitem[{Bachurin et~al.(2010)Bachurin, Weygand and
  Gumbsch}]{bachurin2010dislocation}
\bibinfo{author}{Bachurin, D.V.}, \bibinfo{author}{Weygand, D.},
  \bibinfo{author}{Gumbsch, P.}, \bibinfo{year}{2010}.
\newblock \bibinfo{title}{Dislocation-grain boundary interaction in $\langle
  111\rangle$ textured thin metal films}.
\newblock \bibinfo{journal}{Acta Materialia} \bibinfo{volume}{58},
  \bibinfo{pages}{5232--5241}.
\newblock \DOIprefix\doi{https://doi.org/10.1016/j.jmps.2003.12.007}.
\bibitem[{Balluffi et~al.(1982)Balluffi, Brokman and King}]{balluffi1982csl}
\bibinfo{author}{Balluffi, R.}, \bibinfo{author}{Brokman, A.},
  \bibinfo{author}{King, A.}, \bibinfo{year}{1982}.
\newblock \bibinfo{title}{{CSL/DSC lattice model for general crystalcrystal
  boundaries and their line defects}}.
\newblock \bibinfo{journal}{Acta Metallurgica} \bibinfo{volume}{30},
  \bibinfo{pages}{1453--1470}.
\newblock \DOIprefix\doi{https://doi.org/10.1016/0001-6160(82)90166-3}.
\bibitem[{Bieler et~al.(2014)Bieler, Eisenlohr, Zhang, Phukan and
  Crimp}]{bieler2014grain}
\bibinfo{author}{Bieler, T.}, \bibinfo{author}{Eisenlohr, P.},
  \bibinfo{author}{Zhang, C.}, \bibinfo{author}{Phukan, H.},
  \bibinfo{author}{Crimp, M.}, \bibinfo{year}{2014}.
\newblock \bibinfo{title}{Grain boundaries and interfaces in slip transfer}.
\newblock \bibinfo{journal}{Current Opinion in Solid State and Materials
  Science} \bibinfo{volume}{18}, \bibinfo{pages}{212--226}.
\newblock \DOIprefix\doi{https://doi.org/10.1016/j.cossms.2014.05.003}.
\bibitem[{Cermelli and Gurtin(2002)}]{cermelli2002geometrically}
\bibinfo{author}{Cermelli, P.}, \bibinfo{author}{Gurtin, M.E.},
  \bibinfo{year}{2002}.
\newblock \bibinfo{title}{Geometrically necessary dislocations in viscoplastic
  single crystals and bicrystals undergoing small deformations}.
\newblock \bibinfo{journal}{International Journal of Solids and Structures}
  \bibinfo{volume}{39}, \bibinfo{pages}{6281--6309}.
\newblock \DOIprefix\doi{https://doi.org/10.1016/S0020-7683(02)00491-2}.
\bibitem[{Chang et~al.(2001)Chang, Cai, Bulatov and Yip}]{chang2001dislocation}
\bibinfo{author}{Chang, J.}, \bibinfo{author}{Cai, W.},
  \bibinfo{author}{Bulatov, V.V.}, \bibinfo{author}{Yip, S.},
  \bibinfo{year}{2001}.
\newblock \bibinfo{title}{{Dislocation motion in BCC metals by molecular
  dynamics}}.
\newblock \bibinfo{journal}{Materials Science and Engineering: A}
  \bibinfo{volume}{309}, \bibinfo{pages}{160--163}.
\newblock \DOIprefix\doi{https://doi.org/10.1016/S0921-5093(00)01673-7}.
\bibitem[{Cho et~al.(2020)Cho, Crone, Arsenlis and Aubry}]{cho2020dislocation}
\bibinfo{author}{Cho, J.}, \bibinfo{author}{Crone, J.C.},
  \bibinfo{author}{Arsenlis, A.}, \bibinfo{author}{Aubry, S.},
  \bibinfo{year}{2020}.
\newblock \bibinfo{title}{Dislocation dynamics in polycrystalline materials}.
\newblock \bibinfo{journal}{Modelling and Simulation in Materials Science and
  Engineering} \bibinfo{volume}{28}, \bibinfo{pages}{035009}.
\newblock \DOIprefix\doi{https://doi.org/10.1088/1361-651X/ab6da8}.
\bibitem[{Clark et~al.(1992)Clark, Wagoner, Shen, Lee, Robertson and
  Birnbaum}]{clark1992criteria}
\bibinfo{author}{Clark, W.}, \bibinfo{author}{Wagoner, R.},
  \bibinfo{author}{Shen, Z.}, \bibinfo{author}{Lee, T.},
  \bibinfo{author}{Robertson, I.}, \bibinfo{author}{Birnbaum, H.},
  \bibinfo{year}{1992}.
\newblock \bibinfo{title}{On the criteria for slip transmission across
  interfaces in polycrystals}.
\newblock \bibinfo{journal}{Scripta Metallurgica et Materialia}
  \bibinfo{volume}{26}, \bibinfo{pages}{203--206}.
\newblock \DOIprefix\doi{https://doi.org/10.1016/0956-716X(92)90173-C}.
\bibitem[{Dao et~al.(2006)Dao, Lu, Shen and Suresh}]{dao2006strength}
\bibinfo{author}{Dao, M.}, \bibinfo{author}{Lu, L.}, \bibinfo{author}{Shen,
  Y.}, \bibinfo{author}{Suresh, S.}, \bibinfo{year}{2006}.
\newblock \bibinfo{title}{Strength, strain-rate sensitivity and ductility of
  copper with nanoscale twins}.
\newblock \bibinfo{journal}{Acta Materialia} \bibinfo{volume}{54},
  \bibinfo{pages}{5421--5432}.
\newblock \DOIprefix\doi{https://doi.org/10.1016/j.actamat.2006.06.062}.
\bibitem[{De~Koning et~al.(2003)De~Koning, Kurtz, Bulatov, Henager, Hoagland,
  Cai and Nomura}]{de2003modeling}
\bibinfo{author}{De~Koning, M.}, \bibinfo{author}{Kurtz, R.J.},
  \bibinfo{author}{Bulatov, V.V.}, \bibinfo{author}{Henager, C.H.},
  \bibinfo{author}{Hoagland, R.G.}, \bibinfo{author}{Cai, W.},
  \bibinfo{author}{Nomura, M.}, \bibinfo{year}{2003}.
\newblock \bibinfo{title}{{Modeling of dislocation -- grain boundary
  interactions in FCC metals}}.
\newblock \bibinfo{journal}{Journal of Nuclear Materials}
  \bibinfo{volume}{323}, \bibinfo{pages}{281--289}.
\newblock \DOIprefix\doi{https://doi.org/10.1016/j.jnucmat.2003.08.008}.
\bibitem[{Dewald and Curtin(2006)}]{dewald2006multiscale}
\bibinfo{author}{Dewald, M.}, \bibinfo{author}{Curtin, W.},
  \bibinfo{year}{2006}.
\newblock \bibinfo{title}{{Multiscale modelling of dislocation/grain-boundary
  interactions: I. Edge dislocations impinging on $\Sigma 11$ (113) tilt
  boundary in Al}}.
\newblock \bibinfo{journal}{Modelling and Simulation in Materials Science and
  Engineering} \bibinfo{volume}{15}, \bibinfo{pages}{S193}.
\newblock \DOIprefix\doi{10.1088/0965-0393/15/1/S16}.
\bibitem[{Dewald and Curtin(2007)}]{dewald2007multiscale}
\bibinfo{author}{Dewald, M.}, \bibinfo{author}{Curtin, W.},
  \bibinfo{year}{2007}.
\newblock \bibinfo{title}{Multiscale modelling of dislocation/grain boundary
  interactions. ii. screw dislocations impinging on tilt boundaries in al}.
\newblock \bibinfo{journal}{Philosophical Magazine} \bibinfo{volume}{87},
  \bibinfo{pages}{4615--4641}.
\newblock \DOIprefix\doi{https://doi.org/10.1080/14786430701297590}.
\bibitem[{Dewald and Curtin(2011)}]{dewald2011multiscale}
\bibinfo{author}{Dewald, M.}, \bibinfo{author}{Curtin, W.},
  \bibinfo{year}{2011}.
\newblock \bibinfo{title}{{Multiscale modeling of dislocation/grain-boundary
  interactions: III. $60^\circ$ dislocations impinging on $\Sigma 3$, $\Sigma
  9$ and $\Sigma 11$ tilt boundaries in Al}}.
\newblock \bibinfo{journal}{Modelling and Simulation in Materials Science and
  Engineering} \bibinfo{volume}{19}, \bibinfo{pages}{055002}.
\newblock \DOIprefix\doi{10.1088/0965-0393/19/5/055002}.
\bibitem[{Dunne et~al.(2007)Dunne, Rugg and Walker}]{dunne2007lengthscale}
\bibinfo{author}{Dunne, F.}, \bibinfo{author}{Rugg, D.},
  \bibinfo{author}{Walker, A.}, \bibinfo{year}{2007}.
\newblock \bibinfo{title}{Lengthscale-dependent, elastically anisotropic,
  physically-based hcp crystal plasticity: Application to cold-dwell fatigue in
  ti alloys}.
\newblock \bibinfo{journal}{International Journal of Plasticity}
  \bibinfo{volume}{23}, \bibinfo{pages}{1061--1083}.
\newblock \DOIprefix\doi{https://doi.org/10.1016/j.ijplas.2006.10.013}.
\bibitem[{Elkajbaji and Thibault-Desseaux(1988)}]{elkajbaji1988interactions}
\bibinfo{author}{Elkajbaji, M.}, \bibinfo{author}{Thibault-Desseaux, J.},
  \bibinfo{year}{1988}.
\newblock \bibinfo{title}{{Interactions of deformation-induced dislocations
  with $\Sigma= 9$ (122) grain boundaries in Si studied by HREM}}.
\newblock \bibinfo{journal}{Philosophical Magazine A} \bibinfo{volume}{58},
  \bibinfo{pages}{325--345}.
\newblock \DOIprefix\doi{https://doi.org/10.1080/01418618808209929}.
\bibitem[{Erdle and B{\"o}hlke(2017)}]{erdle2017gradient}
\bibinfo{author}{Erdle, H.}, \bibinfo{author}{B{\"o}hlke, T.},
  \bibinfo{year}{2017}.
\newblock \bibinfo{title}{A gradient crystal plasticity theory for large
  deformations with a discontinuous accumulated plastic slip}.
\newblock \bibinfo{journal}{Computational Mechanics} \bibinfo{volume}{60},
  \bibinfo{pages}{923--942}.
\newblock \DOIprefix\doi{10.1007/s00466-017-1447-7}.
\bibitem[{Evers et~al.(2004)Evers, Brekelmans and Geers}]{evers2004scale}
\bibinfo{author}{Evers, L.}, \bibinfo{author}{Brekelmans, W.},
  \bibinfo{author}{Geers, M.}, \bibinfo{year}{2004}.
\newblock \bibinfo{title}{Scale dependent crystal plasticity framework with
  dislocation density and grain boundary effects}.
\newblock \bibinfo{journal}{International Journal of solids and structures}
  \bibinfo{volume}{41}, \bibinfo{pages}{5209--5230}.
\newblock \DOIprefix\doi{https://doi.org/10.1016/j.ijsolstr.2004.04.021}.
\bibitem[{Fan et~al.(2012)Fan, Li and Huang}]{fan2012toward}
\bibinfo{author}{Fan, H.}, \bibinfo{author}{Li, Z.}, \bibinfo{author}{Huang,
  M.}, \bibinfo{year}{2012}.
\newblock \bibinfo{title}{Toward a further understanding of intermittent
  plastic responses in the compressed single/bicrystalline micropillars}.
\newblock \bibinfo{journal}{Scripta materialia} \bibinfo{volume}{66},
  \bibinfo{pages}{813--816}.
\newblock \DOIprefix\doi{https://doi.org/10.1016/j.scriptamat.2012.02.023}.
\bibitem[{Fan et~al.(2021)Fan, Wang, El-Awady, Raabe and
  Zaiser}]{fan2021strain}
\bibinfo{author}{Fan, H.}, \bibinfo{author}{Wang, Q.},
  \bibinfo{author}{El-Awady, J.A.}, \bibinfo{author}{Raabe, D.},
  \bibinfo{author}{Zaiser, M.}, \bibinfo{year}{2021}.
\newblock \bibinfo{title}{Strain rate dependency of dislocation plasticity}.
\newblock \bibinfo{journal}{Nature Communications} \bibinfo{volume}{12},
  \bibinfo{pages}{1--11}.
\newblock \DOIprefix\doi{https://doi.org/10.1038/s41467-021-21939-1}.
\bibitem[{Fredriksson and Gudmundson(2007)}]{fredriksson2007modelling}
\bibinfo{author}{Fredriksson, P.}, \bibinfo{author}{Gudmundson, P.},
  \bibinfo{year}{2007}.
\newblock \bibinfo{title}{Modelling of the interface between a thin film and a
  substrate within a strain gradient plasticity framework}.
\newblock \bibinfo{journal}{Journal of the Mechanics and Physics of Solids}
  \bibinfo{volume}{55}, \bibinfo{pages}{939--955}.
\newblock \DOIprefix\doi{https://doi.org/10.1016/j.jmps.2006.11.001}.
\bibitem[{Gleiter and Hornbogen(1965)}]{gleiter1965beobachtung}
\bibinfo{author}{Gleiter, H.}, \bibinfo{author}{Hornbogen, E.},
  \bibinfo{year}{1965}.
\newblock \bibinfo{title}{{Beobachtung der Wechselwirkung von Versetzungen mit
  koh{\"a}renten geordneten Zonen (II)}}.
\newblock \bibinfo{journal}{Physica Status Solidi (b)} \bibinfo{volume}{12},
  \bibinfo{pages}{251--264}.
\newblock \DOIprefix\doi{https://doi.org/10.1002/pssb.19650120123}.
\bibitem[{Gottschalk et~al.(2016)Gottschalk, McBride, Reddy, Javili, Wriggers
  and Hirschberger}]{gottschalk2016computational}
\bibinfo{author}{Gottschalk, D.}, \bibinfo{author}{McBride, A.},
  \bibinfo{author}{Reddy, B.}, \bibinfo{author}{Javili, A.},
  \bibinfo{author}{Wriggers, P.}, \bibinfo{author}{Hirschberger, C.},
  \bibinfo{year}{2016}.
\newblock \bibinfo{title}{Computational and theoretical aspects of a
  grain-boundary model that accounts for grain misorientation and
  grain-boundary orientation}.
\newblock \bibinfo{journal}{Computational Materials Science}
  \bibinfo{volume}{111}, \bibinfo{pages}{443--459}.
\newblock \DOIprefix\doi{https://doi.org/10.1016/j.commatsci.2015.09.048}.
\bibitem[{Gurtin(2008)}]{gurtin2008theory}
\bibinfo{author}{Gurtin, M.E.}, \bibinfo{year}{2008}.
\newblock \bibinfo{title}{A theory of grain boundaries that accounts
  automatically for grain misorientation and grain-boundary orientation}.
\newblock \bibinfo{journal}{Journal of the Mechanics and Physics of Solids}
  \bibinfo{volume}{56}, \bibinfo{pages}{640--662}.
\newblock \DOIprefix\doi{https://doi.org/10.1016/j.jmps.2007.05.002}.
\bibitem[{Hall(1951)}]{hall1951deformation}
\bibinfo{author}{Hall, E.O.}, \bibinfo{year}{1951}.
\newblock \bibinfo{title}{{The deformation and ageing of mild steel: III
  discussion of results}}.
\newblock \bibinfo{journal}{Proceedings of the Physical Society. Section B}
  \bibinfo{volume}{64}, \bibinfo{pages}{747}.
\newblock \DOIprefix\doi{10.1088/0370-1301/64/9/303}.
\bibitem[{Han et~al.(2018)Han, Thomas and Srolovitz}]{han2018grain}
\bibinfo{author}{Han, J.}, \bibinfo{author}{Thomas, S.L.},
  \bibinfo{author}{Srolovitz, D.J.}, \bibinfo{year}{2018}.
\newblock \bibinfo{title}{Grain-boundary kinetics: A unified approach}.
\newblock \bibinfo{journal}{Progress in Materials Science}
  \bibinfo{volume}{98}, \bibinfo{pages}{386--476}.
\newblock \DOIprefix\doi{https://doi.org/10.1016/j.pmatsci.2018.05.004}.
\bibitem[{Hochrainer et~al.(2007)Hochrainer, Zaiser and
  Gumbsch}]{hochrainer2007three}
\bibinfo{author}{Hochrainer, T.}, \bibinfo{author}{Zaiser, M.},
  \bibinfo{author}{Gumbsch, P.}, \bibinfo{year}{2007}.
\newblock \bibinfo{title}{A three-dimensional continuum theory of dislocation
  systems: kinematics and mean-field formulation}.
\newblock \bibinfo{journal}{Philosophical Magazine} \bibinfo{volume}{87},
  \bibinfo{pages}{1261--1282}.
\newblock \DOIprefix\doi{https://doi.org/10.1080/14786430600930218}.
\bibitem[{Jiang et~al.(2019)Jiang, Devincre and Monnet}]{jiang2019effects}
\bibinfo{author}{Jiang, M.}, \bibinfo{author}{Devincre, B.},
  \bibinfo{author}{Monnet, G.}, \bibinfo{year}{2019}.
\newblock \bibinfo{title}{Effects of the grain size and shape on the flow
  stress: A dislocation dynamics study}.
\newblock \bibinfo{journal}{International Journal of Plasticity}
  \bibinfo{volume}{113}, \bibinfo{pages}{111--124}.
\newblock \DOIprefix\doi{https://doi.org/10.1016/j.ijplas.2018.09.008}.
\bibitem[{Jiang et~al.(2022)Jiang, Fan, Kruch and Devincre}]{jiang2022grain}
\bibinfo{author}{Jiang, M.}, \bibinfo{author}{Fan, Z.}, \bibinfo{author}{Kruch,
  S.}, \bibinfo{author}{Devincre, B.}, \bibinfo{year}{2022}.
\newblock \bibinfo{title}{Grain size effect of fcc polycrystal: A new cpfem
  approach based on surface geometrically necessary dislocations}.
\newblock \bibinfo{journal}{International Journal of Plasticity}
  \bibinfo{volume}{150}, \bibinfo{pages}{103181}.
\newblock \DOIprefix\doi{https://doi.org/10.1016/j.ijplas.2021.103181}.
\bibitem[{Jin et~al.(2008)Jin, Gumbsch, Albe, Ma, Lu, Gleiter and
  Hahn}]{jin2008interactions}
\bibinfo{author}{Jin, Z.H.}, \bibinfo{author}{Gumbsch, P.},
  \bibinfo{author}{Albe, K.}, \bibinfo{author}{Ma, E.}, \bibinfo{author}{Lu,
  K.}, \bibinfo{author}{Gleiter, H.}, \bibinfo{author}{Hahn, H.},
  \bibinfo{year}{2008}.
\newblock \bibinfo{title}{Interactions between non-screw lattice dislocations
  and coherent twin boundaries in face-centered cubic metals}.
\newblock \bibinfo{journal}{Acta Materialia} \bibinfo{volume}{56},
  \bibinfo{pages}{1126--1135}.
\newblock \DOIprefix\doi{https://doi.org/10.1016/j.actamat.2007.11.020}.
\bibitem[{Jin et~al.(2006)Jin, Gumbsch, Ma, Albe, Lu, Hahn and
  Gleiter}]{jin2006interaction}
\bibinfo{author}{Jin, Z.H.}, \bibinfo{author}{Gumbsch, P.},
  \bibinfo{author}{Ma, E.}, \bibinfo{author}{Albe, K.}, \bibinfo{author}{Lu,
  K.}, \bibinfo{author}{Hahn, H.}, \bibinfo{author}{Gleiter, H.},
  \bibinfo{year}{2006}.
\newblock \bibinfo{title}{The interaction mechanism of screw dislocations with
  coherent twin boundaries in different face-centred cubic metals}.
\newblock \bibinfo{journal}{Scripta Materialia} \bibinfo{volume}{54},
  \bibinfo{pages}{1163--1168}.
\newblock \DOIprefix\doi{https://doi.org/10.1016/j.scriptamat.2005.11.072}.
\bibitem[{Johnston and Gilman(1959)}]{johnston1959dislocation}
\bibinfo{author}{Johnston, W.G.}, \bibinfo{author}{Gilman, J.J.},
  \bibinfo{year}{1959}.
\newblock \bibinfo{title}{Dislocation velocities, dislocation densities, and
  plastic flow in lithium fluoride crystals}.
\newblock \bibinfo{journal}{Journal of Applied Physics} \bibinfo{volume}{30},
  \bibinfo{pages}{129--144}.
\newblock \DOIprefix\doi{https://doi.org/10.1063/1.1735121}.
\bibitem[{Kacher et~al.(2014)Kacher, Eftink, Cui and
  Robertson}]{kacher2014dislocation}
\bibinfo{author}{Kacher, J.}, \bibinfo{author}{Eftink, B.},
  \bibinfo{author}{Cui, B.}, \bibinfo{author}{Robertson, I.},
  \bibinfo{year}{2014}.
\newblock \bibinfo{title}{Dislocation interactions with grain boundaries}.
\newblock \bibinfo{journal}{Current Opinion in Solid State and Materials
  Science} \bibinfo{volume}{18}, \bibinfo{pages}{227--243}.
\newblock \DOIprefix\doi{https://doi.org/10.1016/j.cossms.2014.05.004}.
\bibitem[{Kacher and Robertson(2012)}]{kacher2012quasi}
\bibinfo{author}{Kacher, J.}, \bibinfo{author}{Robertson, I.},
  \bibinfo{year}{2012}.
\newblock \bibinfo{title}{Quasi-four-dimensional analysis of dislocation
  interactions with grain boundaries in 304 stainless steel}.
\newblock \bibinfo{journal}{Acta Materialia} \bibinfo{volume}{60},
  \bibinfo{pages}{6657--6672}.
\newblock \DOIprefix\doi{https://doi.org/10.1016/j.actamat.2012.08.036}.
\bibitem[{Kacher and Robertson(2014)}]{kacher2014situ}
\bibinfo{author}{Kacher, J.}, \bibinfo{author}{Robertson, I.M.},
  \bibinfo{year}{2014}.
\newblock \bibinfo{title}{In situ and tomographic analysis of dislocation/grain
  boundary interactions in $\alpha$-titanium}.
\newblock \bibinfo{journal}{Philosophical Magazine} \bibinfo{volume}{94},
  \bibinfo{pages}{814--829}.
\newblock \DOIprefix\doi{https://doi.org/10.1080/14786435.2013.868942}.
\bibitem[{Kegg et~al.(1973)Kegg, Horton and Silcock}]{kegg1973grain}
\bibinfo{author}{Kegg, G.}, \bibinfo{author}{Horton, C.},
  \bibinfo{author}{Silcock, J.}, \bibinfo{year}{1973}.
\newblock \bibinfo{title}{Grain boundary dislocations in aluminium bicrystals
  after high-temperature deformation}.
\newblock \bibinfo{journal}{Philosophical Magazine} \bibinfo{volume}{27},
  \bibinfo{pages}{1041--1055}.
\newblock \DOIprefix\doi{https://doi.org/10.1080/14786437308225816}.
\bibitem[{King and Smith(1980)}]{king1980effects}
\bibinfo{author}{King, A.}, \bibinfo{author}{Smith, D.}, \bibinfo{year}{1980}.
\newblock \bibinfo{title}{The effects on grain-boundary processes of the steps
  in the boundary plane associated with the cores of grain-boundary
  dislocations}.
\newblock \bibinfo{journal}{Acta Crystallographica Section A}
  \bibinfo{volume}{36}, \bibinfo{pages}{335--343}.
\newblock \DOIprefix\doi{https://doi.org/10.1107/S0567739480000782}.
\bibitem[{Kocks et~al.(1975)Kocks, Argon and Ashby}]{kocks1975progress}
\bibinfo{author}{Kocks, U.F.}, \bibinfo{author}{Argon, A.S.},
  \bibinfo{author}{Ashby, M.F.}, \bibinfo{year}{1975}.
\newblock \bibinfo{title}{Kinetics}, in: \bibinfo{booktitle}{Thermodynamics and
  kinetics of slip}. volume~\bibinfo{volume}{19}. chapter~\bibinfo{chapter}{3},
  pp. \bibinfo{pages}{68--109}.
\bibitem[{Kvashin et~al.(2021)Kvashin, Anento, Terentyev, Bakaev and
  Serra}]{kvashin2021interaction}
\bibinfo{author}{Kvashin, N.}, \bibinfo{author}{Anento, N.},
  \bibinfo{author}{Terentyev, D.}, \bibinfo{author}{Bakaev, A.},
  \bibinfo{author}{Serra, A.}, \bibinfo{year}{2021}.
\newblock \bibinfo{title}{{Interaction of a dislocation pileup with $\{332\}$
  tilt grain boundary in bcc metals studied by MD simulations}}.
\newblock \bibinfo{journal}{Physical Review Materials} \bibinfo{volume}{5},
  \bibinfo{pages}{013605}.
\newblock \DOIprefix\doi{https://doi.org/10.1103/PhysRevMaterials.5.013605}.
\bibitem[{Le(2020)}]{le2020two}
\bibinfo{author}{Le, K.C.}, \bibinfo{year}{2020}.
\newblock \bibinfo{title}{Two universal laws for plastic flows and the
  consistent thermodynamic dislocation theory}.
\newblock \bibinfo{journal}{Mechanics Research Communications}
  \bibinfo{volume}{109}, \bibinfo{pages}{103597}.
\newblock \DOIprefix\doi{https://doi.org/10.1016/j.mechrescom.2020.103597}.
\bibitem[{Lee et~al.(1989)Lee, Robertson and Birnbaum}]{lee1989prediction}
\bibinfo{author}{Lee, T.}, \bibinfo{author}{Robertson, I.},
  \bibinfo{author}{Birnbaum, H.}, \bibinfo{year}{1989}.
\newblock \bibinfo{title}{Prediction of slip transfer mechanisms across grain
  boundaries}.
\newblock \bibinfo{journal}{Scripta metallurgica} \bibinfo{volume}{23},
  \bibinfo{pages}{799--803}.
\newblock \DOIprefix\doi{https://doi.org/10.1016/0036-9748(89)90534-6}.
\bibitem[{Lee et~al.(1990a)Lee, Robertson and Birnbaum}]{lee1990situ}
\bibinfo{author}{Lee, T.}, \bibinfo{author}{Robertson, I.},
  \bibinfo{author}{Birnbaum, H.}, \bibinfo{year}{1990}a.
\newblock \bibinfo{title}{An in situ transmission electron microscope
  deformation study of the slip transfer mechanisms in metals}.
\newblock \bibinfo{journal}{Metallurgical Transactions A} \bibinfo{volume}{21},
  \bibinfo{pages}{2437--2447}.
\newblock \DOIprefix\doi{https://doi.org/10.1007/BF02646988}.
\bibitem[{Lee et~al.(1990b)Lee, Robertson and Birnbaum}]{lee1990tem}
\bibinfo{author}{Lee, T.}, \bibinfo{author}{Robertson, I.},
  \bibinfo{author}{Birnbaum, H.}, \bibinfo{year}{1990}b.
\newblock \bibinfo{title}{{TEM in situ deformation study of the interaction of
  lattice dislocations with grain boundaries in metals}}.
\newblock \bibinfo{journal}{Philosophical Magazine A} \bibinfo{volume}{62},
  \bibinfo{pages}{131--153}.
\newblock \DOIprefix\doi{https://doi.org/10.1080/01418619008244340}.
\bibitem[{LeSar and Capolungo(2020)}]{lesar2020advances}
\bibinfo{author}{LeSar, R.}, \bibinfo{author}{Capolungo, L.},
  \bibinfo{year}{2020}.
\newblock \bibinfo{title}{Advances in discrete dislocation dynamics
  simulations}.
\newblock \bibinfo{journal}{Handbook of Materials Modeling: Methods: Theory and
  Modeling} , \bibinfo{pages}{1079--1110}.
\bibitem[{Leung et~al.(2015)Leung, Leung, Cheng and Ngan}]{leung2015new}
\bibinfo{author}{Leung, H.S.}, \bibinfo{author}{Leung, P.S.S.},
  \bibinfo{author}{Cheng, B.}, \bibinfo{author}{Ngan, A.H.W.},
  \bibinfo{year}{2015}.
\newblock \bibinfo{title}{A new dislocation-density-function dynamics scheme
  for computational crystal plasticity by explicit consideration of dislocation
  elastic interactions}.
\newblock \bibinfo{journal}{International Journal of Plasticity}
  \bibinfo{volume}{67}, \bibinfo{pages}{1--25}.
\newblock \DOIprefix\doi{https://doi.org/10.1016/j.ijplas.2014.09.009}.
\bibitem[{Linne et~al.(2020)Linne, Bieler and Daly}]{linne2020effect}
\bibinfo{author}{Linne, M.A.}, \bibinfo{author}{Bieler, T.R.},
  \bibinfo{author}{Daly, S.}, \bibinfo{year}{2020}.
\newblock \bibinfo{title}{The effect of microstructure on the relationship
  between grain boundary sliding and slip transmission in high purity
  aluminum}.
\newblock \bibinfo{journal}{International Journal of Plasticity}
  \bibinfo{volume}{135}, \bibinfo{pages}{102818}.
\newblock \DOIprefix\doi{https://doi.org/10.1016/j.ijplas.2020.102818}.
\bibitem[{Ma et~al.(2006)Ma, Roters and Raabe}]{ma2006studying}
\bibinfo{author}{Ma, A.}, \bibinfo{author}{Roters, F.}, \bibinfo{author}{Raabe,
  D.}, \bibinfo{year}{2006}.
\newblock \bibinfo{title}{Studying the effect of grain boundaries in
  dislocation density based crystal-plasticity finite element simulations}.
\newblock \bibinfo{journal}{International Journal of Solids and Structures}
  \bibinfo{volume}{43}, \bibinfo{pages}{7287--7303}.
\newblock \DOIprefix\doi{https://doi.org/10.1016/j.ijsolstr.2006.07.006}.
\bibitem[{Martyushev and Seleznev(2006)}]{martyushev2006maximum}
\bibinfo{author}{Martyushev, L.M.}, \bibinfo{author}{Seleznev, V.D.},
  \bibinfo{year}{2006}.
\newblock \bibinfo{title}{Maximum entropy production principle in physics,
  chemistry and biology}.
\newblock \bibinfo{journal}{Physics reports} \bibinfo{volume}{426},
  \bibinfo{pages}{1--45}.
\newblock \DOIprefix\doi{https://doi.org/10.1016/j.physrep.2005.12.001}.
\bibitem[{Mayeur et~al.(2015)Mayeur, Beyerlein, Bronkhorst and
  Mourad}]{mayeur2015incorporating}
\bibinfo{author}{Mayeur, J.}, \bibinfo{author}{Beyerlein, I.},
  \bibinfo{author}{Bronkhorst, C.}, \bibinfo{author}{Mourad, H.},
  \bibinfo{year}{2015}.
\newblock \bibinfo{title}{Incorporating interface affected zones into crystal
  plasticity}.
\newblock \bibinfo{journal}{International journal of plasticity}
  \bibinfo{volume}{65}, \bibinfo{pages}{206--225}.
\newblock \DOIprefix\doi{https://doi.org/10.1016/j.ijplas.2014.08.013}.
\bibitem[{Ng and Ngan(2009)}]{ng2009deformation}
\bibinfo{author}{Ng, K.}, \bibinfo{author}{Ngan, A.H.W.}, \bibinfo{year}{2009}.
\newblock \bibinfo{title}{Deformation of micron-sized aluminium bi-crystal
  pillars}.
\newblock \bibinfo{journal}{Philosophical Magazine} \bibinfo{volume}{89},
  \bibinfo{pages}{3013--3026}.
\newblock \DOIprefix\doi{https://doi.org/10.1080/14786430903164614}.
\bibitem[{Onsager(1931)}]{onsager1931reciprocal}
\bibinfo{author}{Onsager, L.}, \bibinfo{year}{1931}.
\newblock \bibinfo{title}{Reciprocal relations in irreversible processes. i.}
\newblock \bibinfo{journal}{Physical review} \bibinfo{volume}{37},
  \bibinfo{pages}{405}.
\newblock \DOIprefix\doi{https://doi.org/10.1103/PhysRev.37.405}.
\bibitem[{{\"O}zdemir and Yal{\c{c}}inkaya(2014)}]{ozdemir2014modeling}
\bibinfo{author}{{\"O}zdemir, {\.I}.}, \bibinfo{author}{Yal{\c{c}}inkaya, T.},
  \bibinfo{year}{2014}.
\newblock \bibinfo{title}{Modeling of dislocation -- grain boundary
  interactions in a strain gradient crystal plasticity framework}.
\newblock \bibinfo{journal}{Computational Mechanics} \bibinfo{volume}{54},
  \bibinfo{pages}{255--268}.
\newblock \DOIprefix\doi{https://doi.org/10.1007/s00466-014-0982-8}.
\bibitem[{Petch(1953)}]{petch1953cleavage}
\bibinfo{author}{Petch, N.J.}, \bibinfo{year}{1953}.
\newblock \bibinfo{title}{The cleavage strength of polycrystals}.
\newblock \bibinfo{journal}{Journal of the Iron and Steel Institute}
  \bibinfo{volume}{174}, \bibinfo{pages}{25--28}.
\bibitem[{Piao and Le(2022)}]{piao2022thermodynamic}
\bibinfo{author}{Piao, Y.}, \bibinfo{author}{Le, K.C.}, \bibinfo{year}{2022}.
\newblock \bibinfo{title}{Thermodynamic theory of dislocation/grain boundary
  interaction}.
\newblock \bibinfo{journal}{Continuum Mechanics and Thermodynamics}
  \bibinfo{volume}{34}, \bibinfo{pages}{763--780}.
\newblock \DOIprefix\doi{https://doi.org/10.1007/s00161-022-01088-6}.
\bibitem[{Pond and Smith(1977)}]{pond1977absorption}
\bibinfo{author}{Pond, R.C.}, \bibinfo{author}{Smith, D.A.},
  \bibinfo{year}{1977}.
\newblock \bibinfo{title}{On the absorption of dislocations by grain
  boundaries}.
\newblock \bibinfo{journal}{Philosophical Magazine} \bibinfo{volume}{36},
  \bibinfo{pages}{353--366}.
\newblock \DOIprefix\doi{https://doi.org/10.1080/14786437708244939}.
\bibitem[{Quek et~al.(2014)Quek, Wu, Zhang and Srolovitz}]{quek2014polycrystal}
\bibinfo{author}{Quek, S.S.}, \bibinfo{author}{Wu, Z.}, \bibinfo{author}{Zhang,
  Y.W.}, \bibinfo{author}{Srolovitz, D.J.}, \bibinfo{year}{2014}.
\newblock \bibinfo{title}{Polycrystal deformation in a discrete dislocation
  dynamics framework}.
\newblock \bibinfo{journal}{Acta materialia} \bibinfo{volume}{75},
  \bibinfo{pages}{92--105}.
\newblock \DOIprefix\doi{https://doi.org/10.1016/j.actamat.2014.04.063}.
\bibitem[{Roters et~al.(2010)Roters, Eisenlohr, Hantcherli, Tjahjanto, Bieler
  and Raabe}]{roters2010overview}
\bibinfo{author}{Roters, F.}, \bibinfo{author}{Eisenlohr, P.},
  \bibinfo{author}{Hantcherli, L.}, \bibinfo{author}{Tjahjanto, D.D.},
  \bibinfo{author}{Bieler, T.R.}, \bibinfo{author}{Raabe, D.},
  \bibinfo{year}{2010}.
\newblock \bibinfo{title}{Overview of constitutive laws, kinematics,
  homogenization and multiscale methods in crystal plasticity finite-element
  modeling: Theory, experiments, applications}.
\newblock \bibinfo{journal}{Acta Materialia} \bibinfo{volume}{58},
  \bibinfo{pages}{1152--1211}.
\newblock \DOIprefix\doi{https://doi.org/10.1016/j.actamat.2009.10.058}.
\bibitem[{Sangid et~al.(2011)Sangid, Ezaz, Sehitoglu and
  Robertson}]{sangid2011energy}
\bibinfo{author}{Sangid, M.D.}, \bibinfo{author}{Ezaz, T.},
  \bibinfo{author}{Sehitoglu, H.}, \bibinfo{author}{Robertson, I.M.},
  \bibinfo{year}{2011}.
\newblock \bibinfo{title}{Energy of slip transmission and nucleation at grain
  boundaries}.
\newblock \bibinfo{journal}{Acta Materialia} \bibinfo{volume}{59},
  \bibinfo{pages}{283--296}.
\newblock \DOIprefix\doi{https://doi.org/10.1016/j.actamat.2010.09.032}.
\bibitem[{Shen et~al.(1988)Shen, Wagoner and Clark}]{SHEN19883231}
\bibinfo{author}{Shen, Z.}, \bibinfo{author}{Wagoner, R.},
  \bibinfo{author}{Clark, W.}, \bibinfo{year}{1988}.
\newblock \bibinfo{title}{Dislocation and grain boundary interactions in
  metals}.
\newblock \bibinfo{journal}{Acta Metallurgica} \bibinfo{volume}{36},
  \bibinfo{pages}{3231--3242}.
\newblock \DOIprefix\doi{https://doi.org/10.1016/0001-6160(88)90058-2}.
\bibitem[{Stricker et~al.(2016)Stricker, Gagel, Schmitt, Schulz, Weygand and
  Gumbsch}]{stricker2016slip}
\bibinfo{author}{Stricker, M.}, \bibinfo{author}{Gagel, J.},
  \bibinfo{author}{Schmitt, S.}, \bibinfo{author}{Schulz, K.},
  \bibinfo{author}{Weygand, D.}, \bibinfo{author}{Gumbsch, P.},
  \bibinfo{year}{2016}.
\newblock \bibinfo{title}{On slip transmission and grain boundary yielding}.
\newblock \bibinfo{journal}{Meccanica} \bibinfo{volume}{51},
  \bibinfo{pages}{271--278}.
\newblock \DOIprefix\doi{https://doi.org/10.1007/s11012-015-0192-2}.
\bibitem[{Terentyev et~al.(2018)Terentyev, Bakaev, Serra, Pavia, Baker and
  Anento}]{terentyev2018grain}
\bibinfo{author}{Terentyev, D.}, \bibinfo{author}{Bakaev, A.},
  \bibinfo{author}{Serra, A.}, \bibinfo{author}{Pavia, F.},
  \bibinfo{author}{Baker, K.}, \bibinfo{author}{Anento, N.},
  \bibinfo{year}{2018}.
\newblock \bibinfo{title}{Grain boundary mediated plasticity: The role of grain
  boundary atomic structure and thermal activation}.
\newblock \bibinfo{journal}{Scripta Materialia} \bibinfo{volume}{145},
  \bibinfo{pages}{1--4}.
\newblock \DOIprefix\doi{https://doi.org/10.1016/j.scriptamat.2017.10.002}.
\bibitem[{Tsuru et~al.(2016)Tsuru, Shibutani and
  Hirouchi}]{tsuru2016predictive}
\bibinfo{author}{Tsuru, T.}, \bibinfo{author}{Shibutani, Y.},
  \bibinfo{author}{Hirouchi, T.}, \bibinfo{year}{2016}.
\newblock \bibinfo{title}{A predictive model for transferability of plastic
  deformation through grain boundaries}.
\newblock \bibinfo{journal}{AIP Advances} \bibinfo{volume}{6},
  \bibinfo{pages}{015004}.
\newblock \DOIprefix\doi{https://doi.org/10.1063/1.4939819}.
\bibitem[{Tsuru et~al.(2009)Tsuru, Shibutani and Kaji}]{tsuru2009fundamental}
\bibinfo{author}{Tsuru, T.}, \bibinfo{author}{Shibutani, Y.},
  \bibinfo{author}{Kaji, Y.}, \bibinfo{year}{2009}.
\newblock \bibinfo{title}{Fundamental interaction process between pure edge
  dislocation and energetically stable grain boundary}.
\newblock \bibinfo{journal}{Physical Review B} \bibinfo{volume}{79},
  \bibinfo{pages}{012104}.
\newblock \DOIprefix\doi{https://doi.org/10.1103/PhysRevB.79.012104}.
\bibitem[{Van~Beers et~al.(2013)Van~Beers, McShane, Kouznetsova and
  Geers}]{van2013grain}
\bibinfo{author}{Van~Beers, P.}, \bibinfo{author}{McShane, G.},
  \bibinfo{author}{Kouznetsova, V.}, \bibinfo{author}{Geers, M.},
  \bibinfo{year}{2013}.
\newblock \bibinfo{title}{Grain boundary interface mechanics in strain gradient
  crystal plasticity}.
\newblock \bibinfo{journal}{Journal of the Mechanics and Physics of Solids}
  \bibinfo{volume}{61}, \bibinfo{pages}{2659--2679}.
\newblock \DOIprefix\doi{https://doi.org/10.1016/j.jmps.2013.08.011}.
\bibitem[{Wang et~al.(2008)Wang, Hoagland, Hirth and Misra}]{wang2008atomistic}
\bibinfo{author}{Wang, J.}, \bibinfo{author}{Hoagland, R.G.},
  \bibinfo{author}{Hirth, J.P.}, \bibinfo{author}{Misra, A.},
  \bibinfo{year}{2008}.
\newblock \bibinfo{title}{Atomistic modeling of the interaction of glide
  dislocations with “weak” interfaces}.
\newblock \bibinfo{journal}{Acta Materialia} \bibinfo{volume}{56},
  \bibinfo{pages}{5685--5693}.
\newblock \DOIprefix\doi{https://doi.org/10.1016/j.actamat.2008.07.041}.
\bibitem[{Wang et~al.(2022)Wang, Zhang, Zeng, Zhou, He, Liu, Chen, Han,
  Srolovitz, Teng, Guo, Yang, Kong, Ma, Hu, Yin, Huang, Zhang, Zhu and
  Han}]{lihua2022tracking}
\bibinfo{author}{Wang, L.}, \bibinfo{author}{Zhang, Y.}, \bibinfo{author}{Zeng,
  Z.}, \bibinfo{author}{Zhou, H.}, \bibinfo{author}{He, J.},
  \bibinfo{author}{Liu, P.}, \bibinfo{author}{Chen, M.}, \bibinfo{author}{Han,
  J.}, \bibinfo{author}{Srolovitz, D.J.}, \bibinfo{author}{Teng, J.},
  \bibinfo{author}{Guo, Y.}, \bibinfo{author}{Yang, G.}, \bibinfo{author}{Kong,
  D.}, \bibinfo{author}{Ma, E.}, \bibinfo{author}{Hu, Y.},
  \bibinfo{author}{Yin, B.}, \bibinfo{author}{Huang, X.},
  \bibinfo{author}{Zhang, Z.}, \bibinfo{author}{Zhu, T.}, \bibinfo{author}{Han,
  X.}, \bibinfo{year}{2022}.
\newblock \bibinfo{title}{Tracking the sliding of grain boundaries at the
  atomic scale}.
\newblock \bibinfo{journal}{Science} \bibinfo{volume}{375},
  \bibinfo{pages}{1261--1265}.
\newblock \DOIprefix\doi{https://doi.org/10.1126/science.abm2612}.
\bibitem[{Wei et~al.(2003)Wei, Wang, Zhu, Zhou, Ding, Chino and
  Mabuchi}]{wei2003superplasticity}
\bibinfo{author}{Wei, Y.}, \bibinfo{author}{Wang, Q.}, \bibinfo{author}{Zhu,
  Y.}, \bibinfo{author}{Zhou, H.}, \bibinfo{author}{Ding, W.},
  \bibinfo{author}{Chino, Y.}, \bibinfo{author}{Mabuchi, M.},
  \bibinfo{year}{2003}.
\newblock \bibinfo{title}{{Superplasticity and grain boundary sliding in rolled
  AZ91 magnesium alloy at high strain rates}}.
\newblock \bibinfo{journal}{Materials Science and Engineering: A}
  \bibinfo{volume}{360}, \bibinfo{pages}{107--115}.
\newblock \DOIprefix\doi{https://doi.org/10.1016/j.actamat.2008.07.041}.
\bibitem[{Zhang et~al.(2021)Zhang, Lu, Zhang, Tian, Kan and
  Kang}]{zhang2021dislocation}
\bibinfo{author}{Zhang, X.}, \bibinfo{author}{Lu, S.}, \bibinfo{author}{Zhang,
  B.}, \bibinfo{author}{Tian, X.}, \bibinfo{author}{Kan, Q.},
  \bibinfo{author}{Kang, G.}, \bibinfo{year}{2021}.
\newblock \bibinfo{title}{Dislocation--grain boundary interaction-based
  discrete dislocation dynamics modeling and its application to bicrystals with
  different misorientations}.
\newblock \bibinfo{journal}{Acta Materialia} \bibinfo{volume}{202},
  \bibinfo{pages}{88--98}.
\newblock \DOIprefix\doi{https://doi.org/10.1016/j.actamat.2020.10.052}.
\bibitem[{Zhou and LeSar(2012)}]{zhou2012dislocation}
\bibinfo{author}{Zhou, C.}, \bibinfo{author}{LeSar, R.}, \bibinfo{year}{2012}.
\newblock \bibinfo{title}{Dislocation dynamics simulations of plasticity in
  polycrystalline thin films}.
\newblock \bibinfo{journal}{International Journal of Plasticity}
  \bibinfo{volume}{30}, \bibinfo{pages}{185--201}.
\newblock \DOIprefix\doi{https://doi.org/10.1007/978-3-319-44677-6_85}.
\bibitem[{Zhu et~al.(2020)Zhu, Zhao, Deng, An, Song, Mao and Wang}]{ZHU202042}
\bibinfo{author}{Zhu, Q.}, \bibinfo{author}{Zhao, S.}, \bibinfo{author}{Deng,
  C.}, \bibinfo{author}{An, X.}, \bibinfo{author}{Song, K.},
  \bibinfo{author}{Mao, S.}, \bibinfo{author}{Wang, J.}, \bibinfo{year}{2020}.
\newblock \bibinfo{title}{In situ atomistic observation of grain boundary
  migration subjected to defect interaction}.
\newblock \bibinfo{journal}{Acta Materialia} \bibinfo{volume}{199},
  \bibinfo{pages}{42--52}.
\newblock \DOIprefix\doi{https://doi.org/10.1016/j.actamat.2020.08.021}.
\bibitem[{Zhu et~al.(2007)Zhu, Li, Samanta, Kim and
  Suresh}]{zhu2007interfacial}
\bibinfo{author}{Zhu, T.}, \bibinfo{author}{Li, J.}, \bibinfo{author}{Samanta,
  A.}, \bibinfo{author}{Kim, H.G.}, \bibinfo{author}{Suresh, S.},
  \bibinfo{year}{2007}.
\newblock \bibinfo{title}{Interfacial plasticity governs strain rate
  sensitivity and ductility in nanostructured metals}.
\newblock \bibinfo{journal}{Proceedings of the National Academy of Sciences}
  \bibinfo{volume}{104}, \bibinfo{pages}{3031--3036}.
\newblock \DOIprefix\doi{https://doi.org/10.1073/pnas.0611097104}.
\bibitem[{Zhu and Wu(2019)}]{zhu2019perspective}
\bibinfo{author}{Zhu, Y.}, \bibinfo{author}{Wu, X.}, \bibinfo{year}{2019}.
\newblock \bibinfo{title}{Perspective on hetero-deformation induced (hdi)
  hardening and back stress}.
\newblock \bibinfo{journal}{Materials Research Letters} \bibinfo{volume}{7},
  \bibinfo{pages}{393--398}.
\newblock \DOIprefix\doi{https://doi.org/10.1080/21663831.2019.1616331}.
\bibitem[{Zhu and Wu(2023)}]{zhu2022heterostructured}
\bibinfo{author}{Zhu, Y.}, \bibinfo{author}{Wu, X.}, \bibinfo{year}{2023}.
\newblock \bibinfo{title}{Heterostructured materials}.
\newblock \bibinfo{journal}{Progress in Materials Science}
  \bibinfo{volume}{131}, \bibinfo{pages}{101019}.
\newblock \DOIprefix\doi{https://doi.org/10.1016/j.pmatsci.2022.101019}.
\bibitem[{Ziegler(2012)}]{ziegler2012introduction}
\bibinfo{author}{Ziegler, H.}, \bibinfo{year}{2012}.
\newblock \bibinfo{title}{An introduction to thermomechanics}.
\newblock \bibinfo{publisher}{Elsevier}.

\end{thebibliography}
\end{document}